\documentclass[twocolumn, 10pt]{IEEEtran}
\topmargin
-0.8in
\DeclareUnicodeCharacter{FF0C}{ }

\textheight
9.9in

\ifCLASSINFOpdf
\usepackage[pdftex]{graphicx}
\graphicspath{{../pdf/}{../jpeg/}}
\DeclareGraphicsExtensions{.pdf,.jpeg,.png}
\else
\usepackage[dvips]{graphicx}
\graphicspath{{../eps/}}
\DeclareGraphicsExtensions{.eps}
\fi
\usepackage{ifpdf, tabularx}
\usepackage{epstopdf}
\graphicspath{{../}}
\usepackage{threeparttable}

\usepackage{multirow}
\usepackage{float}
\usepackage{amsmath}
\usepackage{amssymb}
\usepackage{array}
\usepackage{mdwmath}
\usepackage{mdwtab}
\usepackage{eqparbox}
\usepackage{nomencl}
\usepackage{cite}
\usepackage{algorithm}
\usepackage{algorithmic}
\usepackage{makeidx}
\usepackage{ifthen}
\usepackage{subfigure}
\usepackage{caption}
\usepackage{pdflscape}
\usepackage{hyperref}
\usepackage{xcolor}

\def\bx{{\mathbf{x}}}

\def\bPg{{\mathbf{P^{g}}}}
\def\WUE{{\mathsf{WUE}}}
\def\EWIF{{\mathsf{EWIF}}}
\def\PUE{{\mathsf{PUE}}}

\newcommand{\beq}{\begin{equation}}
\newcommand{\eeq}{\end{equation}}
\newcommand{\beqn}{\begin{eqnarray}}
\newcommand{\eeqn}{\end{eqnarray}}
\newcommand{\beqno}{\begin{eqnarray*}}
\newcommand{\eeqno}{\end{eqnarray*}}
\newcommand{\bma}{\begin{displaymath}}
\newcommand{\ema}{\end{displaymath}}
\newcommand{\bnu}{\begin{enumerate}}
\newcommand{\enu}{\end{enumerate}}
\newcommand{\bce}{\begin{center}}
\newcommand{\ece}{\end{center}}
\newcommand{\btb}{\begin{tabular}}
\newcommand{\etb}{\end{tabular}}

\newcommand*{\qeda}{\hfill\ensuremath{\blacksquare}}%

\newtheorem{theorem}{Theorem}[section]
\newenvironment{proof}{\noindent\textit{Proof.}}{}

\newtheorem{proposition}[theorem]{Proposition}

\begin{document}
    \title{Green-LLM: Environmentally Aware Workload Allocation for Distributed Inference}
    \author{\IEEEauthorblockN{Jiaming~Cheng,~\IEEEmembership{Student Member,~IEEE}, and~Duong~Tung~Nguyen,~\IEEEmembership{Member,~IEEE}}    \vspace{-0.5cm}
    \thanks{The authors are with Arizona State University, Tempe, AZ, United States.
    Email: \textit{\{jiaming,~duongnt\}@asu.edu.}}
    }

    \maketitle
    \begin{abstract}
        This paper investigates the optimal allocation of large language model (LLM) inference workloads across heterogeneous edge data centers over time. Each data center features on-site renewable generation and faces dynamic electricity prices and spatiotemporal variability in renewable availability. We propose Green-LLM, a lexicographic multi-objective optimization framework that addresses this challenge without requiring manual weight tuning. The proposed model incorporates real-world constraints, including token-dependent processing delay and energy consumption, heterogeneous hardware capabilities, dynamic renewable generation, and spatiotemporal variations in electricity prices and carbon intensity. Unlike existing approaches that optimize individual environmental metrics in isolation, Green-LLM jointly minimizes operational cost, carbon emissions, and delay penalty while enforcing water consumption constraints to ensure both sustainability and quality-of-service requirements. Numerical results demonstrate that Green-LLM achieves significant reductions in carbon emissions and water consumption while maintaining operational costs within 3\% of the minimum and ensuring sub-2-second response latency. These findings show that sustainable LLM inference can be achieved without sacrificing service quality or economic efficiency.
    \end{abstract}

    \printnomenclature

    \allowdisplaybreaks
    \section{Introduction}
    \label{Sec:Intro}

The explosive growth of artificial intelligence (AI) has raised serious environmental concerns in computing. Modern data centers (DCs) consume over 200 TWh annually, with LLM inference alone projected to exceed 100 TWh by 2027~\cite{chien2023reducing}. This massive energy consumption translates to approximately 150 million tons of carbon dioxide emissions and 5 billion cubic meters of water used annually for cooling purposes. As governments worldwide implement carbon pricing mechanisms and water-use restrictions, \textit{sustainable AI computing} has emerged as a critical research field intersecting distributed systems, renewable energy integration, and power grid operation~\cite{shaolei_24}.

The environmental footprint of AI systems extends beyond electricity to encompass both direct and indirect resource usage. Direct impacts include on-site cooling water consumption, characterized by the Water Usage Effectiveness (WUE) metric, while indirect impacts arise from the water and carbon intensity of electricity generation in regional power grids. These environmental costs vary significantly across geographic regions and time periods, creating opportunities for intelligent workload management that existing systems fail to capture.

LLMs involve two distinct operational phases: training and inference. Training is typically a one-time, resource-intensive process lasting weeks or months, whereas inference occurs continuously in real time, directly influencing user experience through response quality and latency. Each inference query comprises input tokens and output tokens, with energy consumption scaling proportionally with the token count~\cite{chien2023reducing}. General-purpose models (e.g., GPT-4) offer versatility but incur high energy costs, while task-specific sub-models (e.g., for translation, summarization, or code generation) can reduce energy consumption by 40--60\% with comparable accuracy. The heterogeneity in model efficiency, hardware capabilities, and resource availability across DCs creates a complex optimization landscape for environmentally sustainable inference.

The geographic distribution of DCs introduces additional optimization opportunities due to spatial and temporal variations in operating conditions. Electricity prices fluctuate significantly between peak and off-peak hours and vary widely across regions due to differences in generation mix, demand patterns, and market structures. Carbon intensity ranges from 50 gCO$_2$/kWh in renewable-rich areas to over 800 gCO$_2$/kWh in coal-dependent regions~\cite{chien2023reducing}. Similarly, water consumption factors exhibit pronounced regional variation depending on cooling technologies and local climate conditions. Renewable energy generation also demonstrates significant temporal-spatial variability, with solar availability following diurnal patterns and wind generation exhibiting stochastic fluctuations.

These observations motivate the following research questions that guide our investigation:
\begin{itemize}
    \item \textbf{RQ1:} \textit{How can LLM inference workloads be optimally allocated across heterogeneous, geographically distributed data centers to jointly minimize carbon emissions, operational costs, and delay penalty while meeting water sustainability and quality-of-service requirements?}
    \item \textbf{RQ2:} \textit{Can multi-objective optimization approaches eliminate the need for manual weight tuning while guaranteeing Pareto-optimal solutions in sustainable LLM inference systems?}
\end{itemize}
Addressing these research questions requires jointly considering: (1) heterogeneous hardware and model compatibilities across data centers, (2) token-dependent LLM inference delay and energy consumption characteristics, (3) spatiotemporal variability in electricity prices, carbon intensity, and renewable energy generation, (4) both direct and indirect water usage constraints, and (5) user-perceived latency requirements that define acceptable service quality. Building upon prior efforts, this paper makes the following contributions: 
\begin{itemize}
    \item \textit{\textbf{Modeling:}} We develop a novel multi-objective optimization model for sustainable distributed LLM inference that jointly minimizes operational cost, carbon emissions, and delay penalty under system constraints, including water consumption limits, token-dependent processing delay and energy consumption, and spatiotemporal electricity prices and renewable availability.
    \item \textit{\textbf{Solution Approach:}} We develop a lexicographic optimization method that eliminates manual weight tuning and guarantees Pareto-optimal solutions. Our formal complexity analysis shows the algorithm runs in polynomial time, making it suitable for real-time deployment.
    \item \textit{\textbf{Real-world Validation:}} Numerical results using real-world data from Microsoft AzureLLM, Google Cloud, electricity markets, and carbon intensity databases demonstrate that Green-LLM significantly reduces carbon emissions and water consumption while maintaining operational costs within 3\% of the minimum and ensuring sub-2-second response latency.
\end{itemize}

\noindent\textbf{Paper Organization:} The remainder of this paper is organized as follows. Section~\ref{sec:related_work} reviews related work. Section~\ref{Sec:ProblemFormulation} presents the system model, problem formulation, and the proposed lexicographic optimization approach. Section~\ref{Sec:NumericalResults} provides numerical results. Section~\ref{sec:conclusion} concludes the paper.

\section{Related Work}
\label{sec:related_work}

This section reviews existing literature on sustainable computing and LLM inference optimization. We organize the discussion into four categories, summarize the comparison in Table~\ref{tab:related_work_comparison}, and position our work.

\noindent \textbf{\textit{1) Latency and SLO-aware LLM Scheduling:}}
SeaLLM~\cite{zhao2025seallm} optimizes resource sharing for latency-sensitive workloads by dynamically partitioning GPU memory across co-located models. In~\cite{slo_llm_inference_serving}, SLO-aware GPU frequency scaling is developed to exploit the distinct power--latency profiles of the prefill and decode phases. SkyLB~\cite{xia2025skylb} introduces locality-aware cross-region load balancing exploiting KV-cache affinity, and SplitLLM~\cite{mudvari2024splitllm} addresses model placement by partitioning models across heterogeneous devices. While these approaches achieve impressive latency improvements, they treat environmental considerations as secondary or ignore them entirely.

\noindent  \textbf{\textit{2) Energy-efficient LLM Inference:}}
In~\cite{wilkinsoffline}, offline energy-optimal serving strategies are developed capturing the relationship between batch size, sequence length, and per-token energy on heterogeneous systems. DynamoLLM in ~\cite{stojkovic2024dynamollm} dynamically adjusts GPU configurations for joint performance and energy efficiency. In~\cite{tian2024greenllm}, energy-aware pruning is introduced that selectively removes redundant attention heads, and~\cite{wallace2025optimization} surveys broader strategies including quantization, distillation, and speculative decoding. However, these methods operate within single DC without considering carbon emissions, water consumption, or geographic workload routing.

\noindent \textbf{\textit{3) Carbon-aware Computing:}}
In~\cite{stojkovic2403towards}, carbon-aware LLM inference is explored by routing workloads based on regional carbon intensity, demonstrating that temporal and spatial shifting can reduce emissions without significant latency degradation. In~\cite{shaolei_24}, geographical load balancing is formulated as a constrained optimization distributing workloads according to regional carbon factors. In~\cite{chien2023reducing}, the carbon impact of generative AI is projected through 2035, identifying geographic shifting as a key mitigation strategy. In~\cite{zhang2024sustainable}, multi-agent reinforcement learning is applied to jointly optimize energy cost and carbon for AIGC workloads across geo-distributed DCs. However, these approaches focus on carbon as the primary metric, overlooking water consumption, and most employ heuristic or learning-based methods without formal optimality guarantees.

\noindent \textbf{\textit{4) Water-aware Data Centers:}}
In~\cite{Li2024_acm}, it is highlighted that AI systems consume substantial water both directly (cooling) and indirectly (electricity generation), with a single GPT-3 training run consuming an estimated 700,000 liters. In~\cite{wue_2024}, comprehensive WUE and EWIF datasets are provided enabling quantitative water-aware research. Concurrent with our work, SLIT~\cite{pasricha2025slit} combines ML prediction with an evolutionary algorithm to generate a Pareto front optimizing latency, carbon, water, and cost for geo-distributed LLM inference. \cite{sust2025deploy} employs MILP to jointly optimize carbon and water under SLO constraints, notably showing that carbon-optimal routing can inadvertently increase water consumption. While these concurrent efforts share our multi-metric vision, they rely on heuristic or weighted-sum methods; our lexicographic approach provides exact polynomial-time optimization without manual weight tuning.

\noindent \textbf{\textit{Positioning of This Work:}}
As summarized in Table~\ref{tab:related_work_comparison}, existing approaches are limited by: (1) single-objective optimization, (2) single DC scope, (3) heuristic methods without optimality guarantees, and (4) static analysis ignoring temporal dynamics. Green-LLM addresses these gaps through a unified multi-objective framework with a lexicographic approach that eliminates manual weight tuning while maintaining polynomial-time complexity.

\begin{table*}[t]
\centering
\caption{Comparison of LLM Inference Optimization Approaches}
\label{tab:related_work_comparison}
\begin{threeparttable}
\begin{tabular}{|l|c|c|c|c|c|c|}
\hline
\textbf{Work} & \textbf{Optimization} & \textbf{Environmental} & \textbf{System} & \textbf{Objective} & \textbf{Infrastructure} & \textbf{Temporal} \\
 & \textbf{Objectives} & \textbf{Factors} & \textbf{Scopes} & \textbf{Method} & \textbf{Model} & \textbf{Dynamics} \\
\hline
\hline
\cite{wilkinsoffline} & Energy (Single-obj.) & None & Single DC & Single-obj. & GPU/CPU hetero. & Static \\
\hline
\cite{stojkovic2024dynamollm} & Energy, Cost (Bi-obj.) & None & Single DC & Weighted sum & Homogeneous & Static \\
\hline
\cite{tian2024greenllm} & Energy (Single-obj.) & None & Edge device & Single-obj. & Homogeneous & Static \\
\hline
\cite{mudvari2024splitllm} & Throughput, Delay (Bi-obj.)& None & Distributed & Heuristic & Hetero. models & Dynamic \\
\hline
\cite{hoffmann2024improving} & Carbon (Single-obj.) & Grid carbon only & Federated & Single-obj. & N/A & Static \\
\hline
\cite{Li2024_acm} & Water (Single-obj.) & Water & Single DC & Single-obj. & N/A & Static \\
\hline
\cite{stojkovic2403towards} & Energy (Single-obj.) & None & Single DC & Heuristic & GPU/CPU hetero. & Static \\
\hline
\cite{pasricha2025slit} & Carbon, Water, Cost, Delay & Carbon + Water & Distributed & Metaheuristic & Hetero. & Dynamic \\
\hline
\cite{sust2025deploy} & Carbon, Water (Bi-obj.) & Carbon + Water & Distributed & Weighted MILP & N/A & Dynamic \\
\hline
\cite{zhang2024sustainable} & Cost, Carbon (Multi-obj.) & Grid carbon & Distributed & Multi-agent RL & Homogeneous & Dynamic \\
\hline
 & \textbf{Carbon, Water,} & \textbf{On-site renewable} & \textbf{Distributed} & \textbf{Lexicographic} & \textbf{Hetero.} & \textbf{Real-time} \\
\textbf{(Ours)} & \textbf{Cost, Delay} & \textbf{+ Water consumption} & \textbf{DCs} & \textbf{Optimization} & \textbf{ Resources} & \textbf{hourly} \\
\hline
\end{tabular}
\begin{tablenotes}
\small
\item[*] Hetero. = Heterogeneous; RL = Reinforcement Learning
\end{tablenotes}
\end{threeparttable}
\end{table*}

\section{System Model and Problem Formulation}
\label{Sec:ProblemFormulation}

\subsection{System Model}
\label{subsec:sys_model} We consider a setting in which an LLM service provider (SP) manages a set of geo-distributed DCs to serve LLM requests from various areas. The SP aims to optimize LLM workload allocation to minimize energy consumption, ensure high service quality, and promote sustainable operations. Let $\mathcal{I}$ and $\mathcal{J}$ denote the sets of areas and DCs, respectively, with indices $i \in \mathcal{I}$ and $j \in \mathcal{J}$. During the inference, each user generates a type-$k$ query where $k \in \mathcal{K}$, and each query type corresponds to an LLM model optimized for a distinct task such as translation, summarization, or code generation.

To efficiently process these queries, the SP employs a pre-trained classifier to automatically categorize incoming prompts based on their content~\cite{hoffmann2024improving}. Once classified, a type-$k$ query originating from area $i$ is routed to an appropriate DC $j$ hosting the corresponding LLM sub-model, and the response is returned to the source area. Let $a_{j,k}$ be a binary indicator equal to $1$ if DC $j$ hosts the LLM sub-model required to process type-$k$ queries. Let $\lambda_{i,k}^{t}$ denote the number of type-$k$ queries generated in area $i$ at time $t$, and define $x_{i,j,k}^{t} \in [0,1]$ as the fraction of these queries processed by DC $j$. Each type-$k$ query has average input and output token lengths $h_k$ and $f_k$, respectively. The energy and water consumption required to serve a query are proportional to the input and output token counts~\cite{wilkinsoffline}.

To process inference workloads, DCs consume computing resources such as RAM, GPU, and CPU. Let $r \in \mathcal{R}$ indicate computing resource type and $C_{j}^{r}$ represent the capacity of resource type $r$ at DC $j$. Each DC $j$ may have on-site renewable energy sources, such as wind turbines and solar panels. Let $P_{j,t}^{\sf w}$ denote the total renewable energy generated at DC $j$ at time $t$. The SP can also procure electricity from the main grid. To reduce both electricity cost and environmental impacts, the SP should prioritize allocating inference workload to DCs with greater renewable energy availability or lower electricity prices. Given the spatial and temporal variability in electricity prices and renewable generation, workload allocation decisions must be dynamically adjusted over time.

\subsection{Problem Formulation}
\noindent 
In the following, we introduce cost objectives, followed by physical constraints considered in the proposed model.

\noindent
\textit{\textbf{1) Power procurement cost:}} To power the AI DCs, the SP can procure electricity from the main grid, which incurs a cost:
\begin{align}
        \mathcal{C}_{1}(\bPg) = \sum_{j,t}c_{j}^{t} P_{j,t}^{\sf g},
\end{align}
where $P_{j,t}^{\sf g}$ represents the total procured electricity from the grid and $c_{j}^{t}$ is the electricity price at DC $j$ at time $t$. 

\noindent
\textit{\textbf{2) Carbon emission cost:}}  
Let $\theta_{j}^{t}$ denote the carbon intensity, i.e., the emission factor per unit of electricity procured from the grid at DC $j$ at time $t$. The resulting carbon emission at DC $j$ at time $t$ is $l_{j}^{t}= \theta_{j}^t P_{j,t}^{\sf g}, \forall j,t$. Let $\delta_{j}$ represent the carbon price (e.g., via carbon tax or cap-and-trade) at location $j$. The total carbon emission cost is given by~\cite{ICNC_24}: 
\begin{align}
    \mathcal{C}_{2}(\bPg) = \sum_{j,t}\delta_{j} l_{j}^{t}= \sum_{j,t} \delta_{j} \theta_{j}^{t} P_{j,t}^{\sf g}.    
\end{align}

\noindent  \textit{\textbf{3) Delay penalty:}} We consider transmission delay, propagation delay, and processing delay. Let $\beta_{i,k}^{t}$ denote the average size of tokens (in bits) at time $t$, and $B_{i,j}$ denote the available bandwidth on the link between $i$ and $j$. The average transmission delay for query type-$k$ from area $i$ is:
\begin{align}
    D_{i,k,t}^{\sf tran}= \sum_{j}\frac{ \beta_{i,k}^{t} (h_{k}+ f_{k})  x_{i,j,k}^{t} }{B_{i,j}}, ~ \forall i,k,t.
\end{align}
Let $d_{i,j}$ denote the network delay between $i$ and $j$.
The average propagation delay for query type-$k$ from area $i$ is:
\begin{align}
        D_{i,k,t}^{\sf prop}= \sum_{j}d_{i,j} x_{i,j,k}^{t}, ~ \forall i,k,t.
\end{align}
Let $v_{j,k}$ be the average processing delay per unit token of query type-$k$ at DC $j$. The average processing delay for query type-$k$ from area $i$ is:
\begin{align}
    D_{i,k,t}^{\sf proc}= \sum_{j}v_{j,k} (h_k + f_k) x_{i,j,k}^{t}, ~ \forall i,k,t.
\end{align}
Then, the total delay cost can be expressed as:
\begin{align}
    \mathcal{C}_{3}(\bx) = \sum_{i,k,t} \rho^{k} \Big( D_{i,k,t}^{\sf tran} + D_{i,k,t}^{\sf prop} + D_{i,k,t}^{\sf proc} \Big),
\end{align}
where $\rho^k$ is the unit delay penalty.
\noindent
The underlying problem can be cast as a multi-objective linear program (MO-LP):   
\begin{subequations}
        \label{Model:M0}
        \begin{align}
        \textbf{M0:} ~ & \min_{\bPg, \bx} \mathcal{C}_{1}(\bPg);~ \min_{\bPg, \bx}  \mathcal{C}_{2}(\bPg);~ \min_{\bPg, \bx} \mathcal{C}_{3}(\bx) \\
         & \textit{s.t.}~~  P_{j,t}^{\sf d}= P_{j,t}^{\sf g}+ P_{j,t}^{\sf w}, ~\forall j,t \label{constrs:power_balance}\\
         & P_{j,t}^{\sf c} = \sum_{i} \big( \tau_{k}^{\sf in} h_{k} + \tau_{k}^{\sf out} f_{k} \big) \lambda_{i,k}^{t} x_{i,j,k}^{t}, ~\forall j,t  \label{constr:P_compute} \\
         &  P_{j,t}^{\sf d} = \PUE_j P_{j,t}^{\sf c}, ~~ \forall j, t    \label{constr:P_tot_demand} \\
         &  0 \leq P_{j,t}^{\sf g}\leq P_{j,t}^{\sf max}, ~ \forall j,t \label{constr:grid_cap} \\
         &  W_{j,t} = \bigg( \frac{\WUE_{j,t}}{\PUE_{j}} + \EWIF_{j,t} \bigg)  P_{j,t}^{\sf d}, ~~ \forall j,t \label{constr:unit_water}\\
         & \sum_{j,t}W_{j,t}\leq Z \label{constr:water_budget}\\
         & \sum_{j}x_{i,j,k}^{t}= 1;~ 0 \leq x_{i,j,k}^t \leq a_{j,k} ~ \forall i, j, k, t \label{constr:supply-demand} \\
         &  \sum_{i,k}\alpha_{k,r} (f_{k}+h_{k}) \lambda_{i,k}^{t}x_{i,j,k}^{t}\leq C_{j}^{r}, ~ \forall j,r,t \label{Constr:Multi_Cap} \\
         &  D_{i,k,t}^{\sf tran}+ D_{i,k,t}^{\sf prop}+ D_{i,k,t}^{\sf proc}\leq \Delta_{i,k}, \forall i,k,t. \label{constr:avg_delay}
        \end{align}
\end{subequations}

Constraint (\ref{constrs:power_balance}) enforces power balance, where $P_{j,t}^{\sf d}$, $P_{j,t}^{\sf g}$, and $P_{j,t}^{\sf w}$ denote total demand, grid procurement, and renewable generation, respectively. Let $\tau_{k} = (\tau_{k}^{\sf in}, \tau_{k}^{\sf out})$ be energy-consumption coefficients per input/output token for query type $k$. Constraint (\ref{constr:P_compute}) models computation energy as linear in token count~\cite{wilkinsoffline}, while (\ref{constr:P_tot_demand}) relates total demand to computation via the Power Usage Effectiveness (PUE) metric~\cite{Huang15}. Constraint (\ref{constr:grid_cap}) limits grid procurement to capacity $P_{j,t}^{\sf max}$.

Departing from prior work, we incorporate both direct and indirect water consumption. Direct use is captured by Water Usage Effectiveness (WUE), while indirect use from electricity generation is characterized by the Electricity–Water Intensity Factor (EWIF)~\cite{Li2024_acm,wue_2024}. Constraints (\ref{constr:unit_water})--(\ref{constr:water_budget}) enforce water usage limits. Let $\alpha_{k,r}$ denote resource $r$ required per type-$k$ query; constraint (\ref{Constr:Multi_Cap}) ensures sufficient computing capacity. Constraints (\ref{constr:supply-demand}) guarantee full service coverage, and (\ref{constr:avg_delay}) imposes delay threshold $\Delta_{i,k}$. Note that the continuous relaxation $x_{i,j,k}^{t} \in [0,1]$ is justified by the large number of queries per time period, where fractional routing corresponds to probabilistic load splitting across DCs.
\subsection{Weighted-Sum Approach}
\label{subsec:weighted_sum}

A common method to solve the MO-LP (\ref{Model:M0}) is to assign different weights to each cost component, yielding a single-objective formulation:
    \begin{subequations}
        \label{Model:weighted}
        \begin{align}
        \min_{\bPg,\bx} &  ~~ \sigma_{e} \mathcal{C}_{1}(\bPg) + \sigma_{c} \mathcal{C}_{2}(\bPg) + \sigma_{d} \mathcal{C}_{3}(\bx) \\
        & \textit{s.t.}~~ \sigma_{e} + \sigma_{c} + \sigma_{d} = 1\\
        &  (\ref{constrs:power_balance}) - (\ref{constr:avg_delay}).
        \end{align}
    \end{subequations}
Since (\ref{Model:weighted}) is a linear program with $n$ decision variables and $m$ constraints, it can be solved in polynomial time with complexity $O((n+m)^3)$ using interior-point methods~\cite{ge2025interior}.

However, determining appropriate weights $(\sigma_e, \sigma_c, \sigma_d)$ is challenging in practice. Weight tuning typically requires extensive experimentation, as the objectives have heterogeneous units (dollars, tons of CO$_2$, seconds) and vastly different scales. Grid search over the weight simplex is computationally expensive: exploring $W$ weight combinations requires solving $W$ separate LPs, yielding total complexity $O(W(n+m)^3)$. In practice, $W \geq 100$ samples may be needed to adequately cover the Pareto frontier, making real-time weight selection impractical. Moreover, the mapping from weights to Pareto outcomes is non-intuitive—small weight changes can induce large shifts in the solution, complicating interpretability for operators.

\subsection{Lexicographic Optimization Approach}
\label{subsec:lexicographic}

To circumvent the challenges of manual weight tuning, we adopt a lexicographic optimization approach that establishes a strict priority ordering among objectives. This method is particularly well-suited for sustainable LLM inference where clear operational priorities exist: operators typically prioritize cost control (to ensure economic viability), then carbon reduction (to meet sustainability commitments), and finally latency optimization (to enhance user experience). Unlike weighted-sum methods that require continuous weight calibration, lexicographic optimization requires only a discrete ranking of objectives—a decision that aligns naturally with priorities and regulatory requirements.

The lexicographic approach solves a sequence of LPs, where each stage optimizes one objective while constraining all higher-priority objectives to remain within a tolerance $\epsilon$ of their optimal values. The main steps are summarized in \textbf{Algorithm}~\ref{alg:lexi}. Specifically, after solving the LP corresponding to the highest-priority objective, its optimal value is recorded and used as a constraint in subsequent optimization phases. This iterative process continues until all objectives are addressed.

\begin{algorithm}
\caption{Lexicographic Optimization for Green-LLM}
\begin{algorithmic}[1]
\REQUIRE Tolerance $\epsilon$; cost parameters; priority list $\mathcal{O} = [o_1, o_2, o_3]$, where $o_\ell \in \{C_{\text{energy}}, C_{\text{carbon}}, C_{\text{delay}}\}$

\STATE Initialize \texttt{phase\_results} $\leftarrow [\ ]$
\STATE Initialize \texttt{optimal\_values} $\leftarrow \{\}$

\FOR{$\ell = 1$ to $3$}
    \STATE $o \leftarrow o_\ell$ \COMMENT{Current priority objective}
    \STATE Define model $\mathcal{M}_\ell: \min C_o~ \textit{s.t} ~(\ref{constrs:power_balance}) - (\ref{constr:avg_delay})$
    \FOR{each $o' \in \{o_1, \dots, o_{\ell-1}\}$}
        \STATE Add $C_{o'} \leq (1 + \epsilon) \cdot \texttt{optimal\_values}[o']$
    \ENDFOR
    \STATE Update \texttt{optimal\_values}$[o] \leftarrow C_o$
    \STATE Append results to \texttt{phase\_results}
\ENDFOR

\RETURN Final decisions $(\bx, \bPg)$ and \texttt{phase\_results}
\end{algorithmic}
\label{alg:lexi}
\end{algorithm}

For clarity, we present the explicit LP formulations solved at each stage of Algorithm~\ref{alg:lexi}, assuming the priority order: energy cost $\succ$ carbon cost $\succ$ delay penalty. The algorithm proceeds sequentially through these stages without iteration or backtracking.

\smallskip
\noindent \textit{\textbf{Stage 1 - Energy Cost Minimization:}} The first stage minimizes the electricity procurement cost without any additional constraints beyond the original system constraints:
\begin{subequations}
\label{Model:stage1}
\begin{align}
\mathcal{M}_1: \quad & \min_{\bPg, \bx} \quad \mathcal{C}_{1}(\bPg) = \sum_{j,t}c_{j}^{t} P_{j,t}^{\sf g} \\
& \textit{s.t.} \quad (\ref{constrs:power_balance}) - (\ref{constr:avg_delay}).
\end{align}
\end{subequations}
Let $\mathcal{C}_1^* = \mathcal{C}_1(\bPg^{(1)})$ denote the optimal energy cost obtained from $\mathcal{M}_1$. This value establishes the baseline for cost-efficient operation.

\smallskip
\noindent \textit{\textbf{Stage 2 - Carbon Cost Minimization:}} The second stage minimizes carbon emissions while ensuring that the energy cost does not exceed the optimal value from Stage 1 by more than a factor of $(1+\epsilon)$:
\begin{subequations}
\label{Model:stage2}
\begin{align}
\!\! \mathcal{M}_2: ~ & \min_{\bPg, \bx} \quad \mathcal{C}_{2}(\bPg) = \sum_{j,t} \delta_{j} \theta_{j}^{t} P_{j,t}^{\sf g} \\
& \textit{s.t.} \quad (\ref{constrs:power_balance}) - (\ref{constr:avg_delay}) \\
& \qquad~ \mathcal{C}_{1}(\bPg) \leq (1+\epsilon) \mathcal{C}_1^*. \label{constr:stage2_energy}
\end{align}
\end{subequations}
Constraint (\ref{constr:stage2_energy}) preserves near-optimality of energy cost while allowing flexibility for carbon reduction. Let $\mathcal{C}_2^*$ denote the optimal carbon cost from $\mathcal{M}_2$.

\smallskip
\noindent \textit{\textbf{Stage 3 - Delay Penalty Minimization:}} The final stage minimizes the delay penalty while preserving the near-optimality of both energy cost and carbon emissions:
\begin{subequations}
\label{Model:stage3}
\begin{align}
\!\! \mathcal{M}_3: & \min_{\bPg, \bx} ~ \mathcal{C}_{3}(\bx) = \sum_{i,k,t} \rho^{k} \Big( D_{i,k,t}^{\sf tran} \!+\! D_{i,k,t}^{\sf prop} \!+\! D_{i,k,t}^{\sf proc} \Big) \\
& \textit{s.t.} \quad (\ref{constrs:power_balance}) - (\ref{constr:avg_delay}) \\
& \qquad~ \mathcal{C}_{1}(\bPg) \leq (1+\epsilon) \mathcal{C}_1^* \label{constr:stage3_energy}\\
& \qquad~ \mathcal{C}_{2}(\bPg) \leq (1+\epsilon) \mathcal{C}_2^*. \label{constr:stage3_carbon}
\end{align}
\end{subequations}

\noindent Upon completion of Stage 3, the algorithm terminates and returns the final allocation decisions $(\bx^*, \bPg^*)$. This solution is lexicographically optimal: it achieves the minimum energy cost, then the minimum carbon cost subject to near-optimal energy cost, and finally the minimum delay penalty subject to near-optimal values of both higher-priority objectives. No further iterations are required—the algorithm makes a single sequential pass through all $L$ objectives.

\subsection{Complexity Analysis}

We analyze the computational complexity of the proposed lexicographic optimization algorithm. Let $L$ denote the number of objectives (distinct from $\mathcal{K}$, which denotes query types).

\begin{proposition}[Computational Complexity]
\label{prop:complexity}
For a multi-objective linear program with $n$ decision variables, $m$ constraints, and $L$ objectives, Algorithm~\ref{alg:lexi} terminates in polynomial time with worst-case complexity $O(L(n+m)^3)$.
\end{proposition}

\begin{proof}
Let $\mathcal{M}_\ell$ denote the LP solved at stage $\ell \in \{1, \ldots, L\}$. At stage $\ell = 1$, the algorithm solves $\mathcal{M}_1: \min C_{o_1}$ subject to the original $m$ constraints from (\ref{constrs:power_balance})--(\ref{constr:avg_delay}). For each subsequent stage $\ell > 1$, the algorithm adds $\ell - 1$ near-optimality constraints of the form $C_{o'} \leq (1+\epsilon) \cdot C_{o'}^*$ for all previously optimized objectives $o' \in \{o_1, \ldots, o_{\ell-1}\}$. Thus, $\mathcal{M}_\ell$ contains $m + \ell - 1$ constraints.

Each LP $\mathcal{M}_\ell$ can be solved using an interior-point method with worst-case complexity $O((n + m + \ell - 1)^3)$~\cite{ge2025interior}. Since $\ell \leq L$ and $L$ is constant, we have $m + \ell - 1 = O(m)$. Therefore, solving each $\mathcal{M}_\ell$ requires $O((n+m)^3)$ operations. The algorithm executes exactly $L$ stages, yielding a total complexity of $\sum_{\ell=1}^{L} O((n+m)^3) = O(L(n+m)^3)$. For our three-objective formulation ($L=3$), the complexity simplifies to $O((n+m)^3)$. \qeda
\end{proof}

\noindent\textbf{Remark 1 (Comparison with Weighted-Sum):} The weighted-sum approach requires sampling $W$ weight vectors to explore the Pareto frontier, with total complexity $O(W(n+m)^3)$. In practice, $W$ can be large (e.g., $W \geq 100$) to adequately cover the continuous weight space. In contrast, lexicographic optimization requires exactly $L$ LP solves, offering computational advantages when strict priority ordering among objectives is acceptable.

\noindent\textbf{Remark 2 (Pareto Optimality):} The lexicographic solution with tolerance $\epsilon$ yields an $\epsilon$-Pareto optimal solution: no feasible solution can improve any objective by more than factor $(1+\epsilon)$ without degrading a higher-priority objective.

\section{Numerical Results}
\label{Sec:NumericalResults}
\subsection{Simulation Setup}
\subsubsection{Network topology}
We consider a network system comprising $9$ DCs ($J = 9$) and $9$ areas ($I = 9$). The network topology is based on the locations of Google Cloud DCs. Network delays between any two locations are obtained using the global ping dataset\footnote{\url{https://wondernetwork.com/pings}}, while electricity prices are extracted from GridStatus\footnote{\url{https://www.gridstatus.io/}}. Carbon intensity values $(\theta_j^{t})$ are derived from the Google Cloud dataset\footnote{\url{https://cloud.google.com/sustainability/region-carbon}}. Regional carbon taxes are generated by scaling a base carbon tax of $\$50\mathrm{/tCO}_2$ by a region-specific factor. The sensitivity of key parameters, including carbon intensity, renewable availability, electricity price, and water budget, is evaluated in Section~\ref{Sec:NumericalResults}. In the \textit{default setting}, the renewable energy generation $P_{j,t}^{\sf w}$ at each DC is simulated based on wind speeds drawn from a Weibull distribution with shape $\kappa = 2$ and scale $c_w = 7$, then scaled to the range $[500,1000]$ KW to reflect a typical mid-scale wind turbine output. PUE values are taken from publicly available Google Cloud statistics\footnote{\url{https://datacenters.google/efficiency/}}. The \textit{WUE} and \textit{EWIF} values are also obtained from the public sources~\cite{wue_2024}. We consider $R=4$ different types of physical resources, with their capacities ($C_j^r$) based on specifications of selected Google Cloud DCs\footnote{\url{https://cloud.google.com/compute/docs/machine-resource}}.

\subsubsection{LLM inference workload}
We generate the time-varying query demand $\lambda_{i,k}^{t}$ using a base hourly demand signal that captures 24-hour usage patterns (i.e., $T = 24$), scaled by a region-specific population multiplier to reflect spatial heterogeneity. We consider five types of task-specific queries (i.e., $K =5$). 
Each task type $k$ is associated with an average number of input tokens $h_k \!=\! \{120,500,70,80,30\}$ and output tokens $f_k \!=\! \{150,350,200,600,20\}$. Each task type is weighted by a query-type popularity factor to reflect how commonly each type appears in practice (e.g., $a^{\text{pop}}_k = [2.5,1.5,1.3,0.8,0.6]$). 
To model the temporal demand fluctuations, we modify the base demand by using a ``\textit{time-of-use}" demand to differentiate peak and off-peak periods~\cite{wilkinsoffline}. Specifically, for each area and query type, demand during peak hours (2:00 pm to 8:00 pm) is drawn from a uniform interval $[900,1000]$, while the off-peak demand is randomly generated from the interval $[500,600]$ based on statistical characteristics in Azure LLM public dataset \footnote{\url{https://github.com/Azure/AzurePublicDataset/}}. For clarity, we use Pacific Time Zone (PST) as the reference.

\subsubsection{Other parameters} 
The processing delay for each type-$k$ query is generated based on task complexity, i.e., $v_{j,k} = [10^{-3},0.002,10^{-2},0.02,0.03]$\,ms/token. For simplicity, the average delay threshold $\Delta_{i,k}$ is set to $5$\,s, $\forall i,k$. \textit{All the experiments are conducted in Python using Gurobipy\footnote{https://pypi.org/project/gurobipy/} on a desktop with an Intel Core i7-11700KF CPU and $32$GB of RAM.} The source code, dataset, and models are available at: \url{https://github.com/JJmingcc/Green_LLM}. 

\subsection{Performance Evaluation}
We first evaluate the weighted-sum formulation of \textbf{M0} with $(\sigma_e,\sigma_c,\sigma_d) = (\frac{1}{3},\frac{1}{3},\frac{1}{3})$ by comparing it against the following benchmark schemes:
\begin{itemize}
    \item \textbf{M1} (Cost-Saving Model~\cite{wilkinsoffline}): Prioritizes allocating LLM workloads to DCs with the lowest electricity prices, where $(\sigma_e,\sigma_c,\sigma_d) = (1,0,0)$.
    \item \textbf{M2} (Carbon-Saving Model~\cite{ICNC_24}): Prioritizes allocating LLM workloads to DCs with the lowest carbon emissions, where $(\sigma_e,\sigma_c,\sigma_d) = (0,1,0)$.
\end{itemize}
Models \textbf{M1}–\textbf{M2} optimize individual objectives (cost and carbon, respectively). We also consider six heuristics: \textbf{H1} (greedy cost-min~\cite{chien2023reducing}), \textbf{H2} (renewable-first), \textbf{H3} (delay-aware~\cite{zhao2025seallm}), \textbf{H4} (carbon-aware~\cite{stojkovic2403towards}), \textbf{H5} (round-robin), and \textbf{H6} (nearest-DC).
We then evaluate the performance of all models in terms of the following metrics.
\begin{itemize}
    \item \textit{\textbf{Total cost:}}  $ \mathcal{C}^{\sf a} = \mathcal{C}_1 (\bPg) +  \mathcal{C}_2 (\bPg) + \mathcal{C}_3 (\bx) $
    \item \textit{\textbf{Carbon emission:}} 
    $ \mathcal{CO}_2 = \mathbf{\theta}^{\top}\bPg. $
    \item \textit{\textbf{Delay penalty:}}  $\mathcal{C}_3 =\sum_{i,k,t} \rho^{k} \Big( D_{i,k,t}^{\sf tran} + D_{i,k,t}^{\sf prop} + D_{i,k,t}^{\sf proc} \Big)$,
\end{itemize}
where decisions $(\bPg,\bx)$ are obtained from above models. In the sensitivity analyses below, we use scaling factors $\Psi_\theta$, $\Psi_w$, $\Psi_\tau$, $\Psi_\rho$, and $\Psi_Z$ to vary the baseline carbon intensity, renewable energy, token size, delay penalty, and water budget, respectively.

\begin{figure}[t]
    \centering
    \includegraphics[width=0.5\textwidth,height=0.145\textheight]{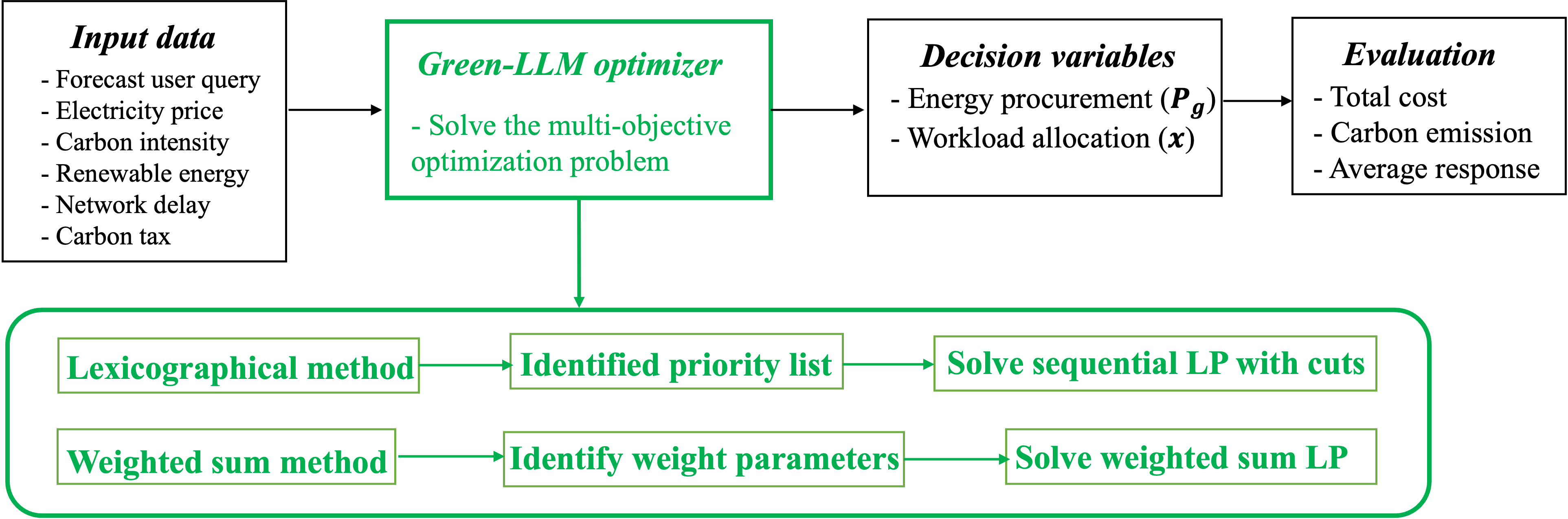}
    \caption{System workflow of the Green-LLM framework.}
    \label{fig:workflow}
\end{figure}

Fig.~\ref{fig:workflow} illustrates the overall workflow of the proposed Green-LLM framework. The system operates in discrete time intervals and consists of four main stages: (1) \textit{Input stage}: retrieve real-time grid data (electricity prices, carbon intensity) from electricity markets (e.g., CAISO, PJM, ERCOT) and generate workload forecasts; (2) \textit{Optimization stage}: solve the multi-objective optimization problem to obtain allocation $\bx$ and power procurement $\bPg$; (3) \textit{Deployment stage}: translate decisions into routing weights or traffic-splitting policies for load balancers; and (4) \textit{Evaluation stage}: track operational metrics for monitoring and continuous improvement.

\vspace{-0.2cm}
\begin{figure}[h!]
\centering
		\subfigure[Total cost (varying $\theta$)]{
	     \includegraphics[width=0.23\textwidth,height=0.125\textheight]{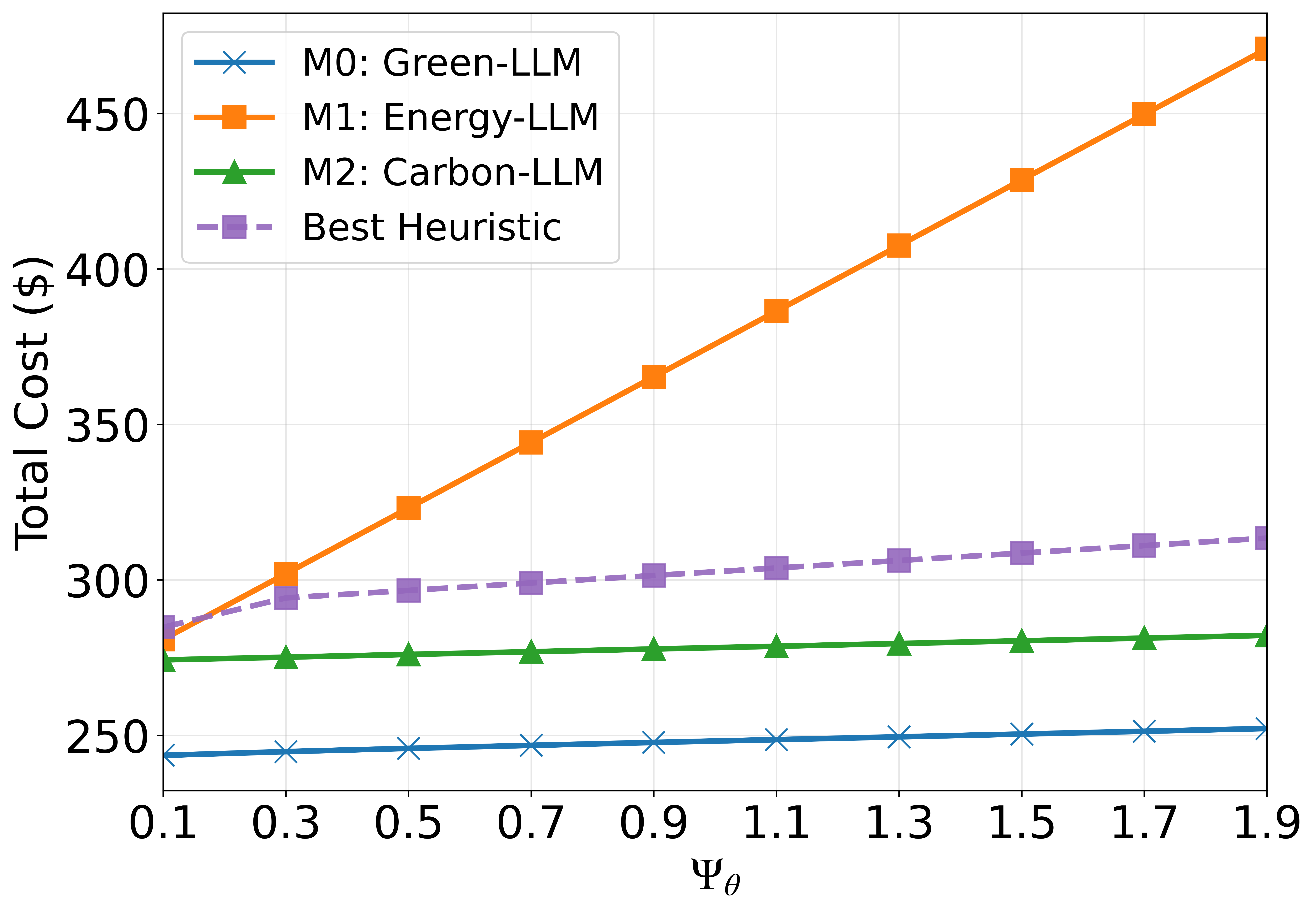}
	     \label{fig:Theta_cost}
	}  \hspace*{-1em} 
	     \subfigure[Total cost (varying $P_w$)]{
	     \includegraphics[width=0.23\textwidth,height=0.125\textheight]{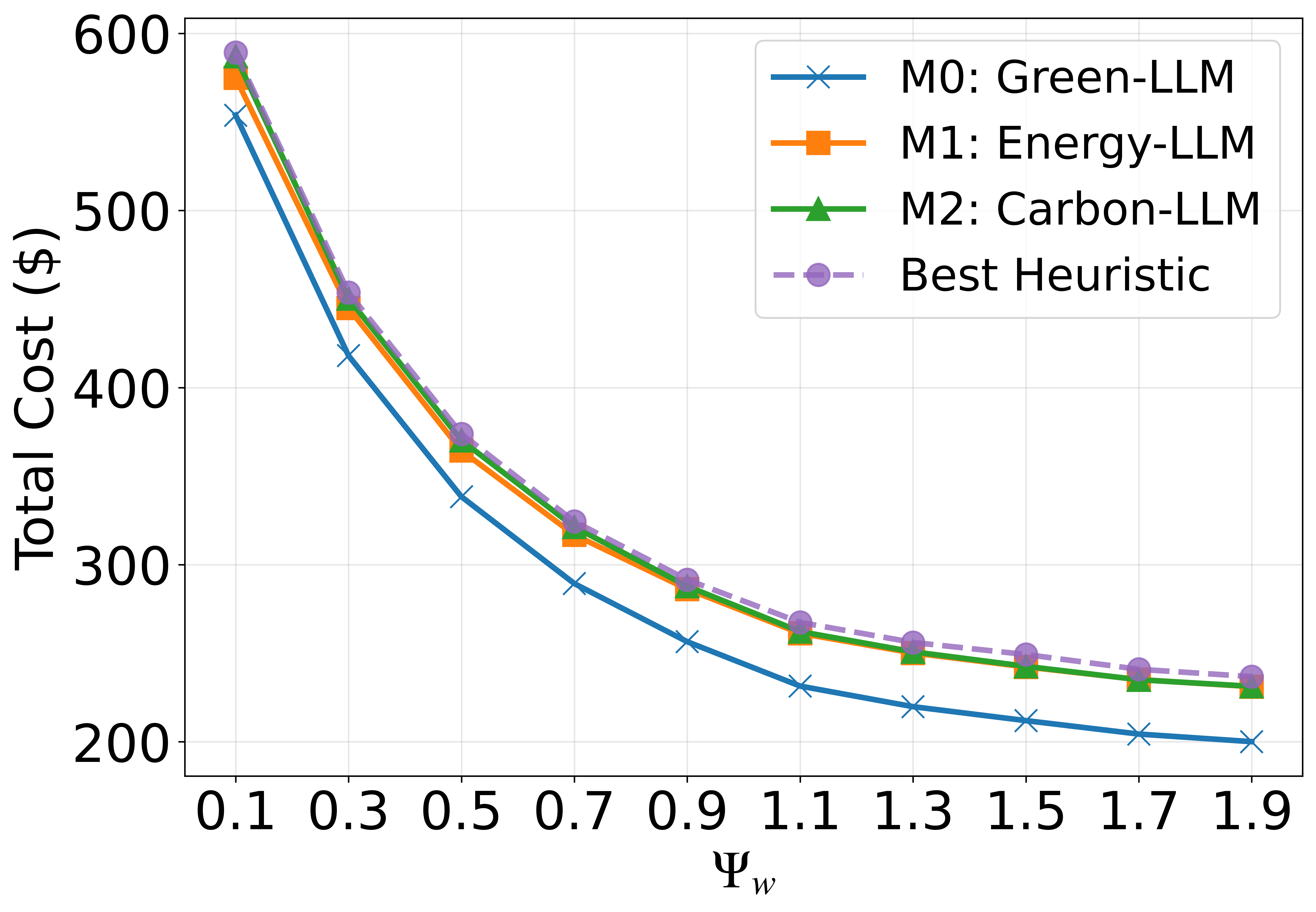}
	     \label{fig:Pw_cost}
	}\vspace{-0.1cm}
    \subfigure[Total cost (t): $\Psi_\theta = 0.7$]{
	     \includegraphics[width=0.23\textwidth,height=0.125\textheight]{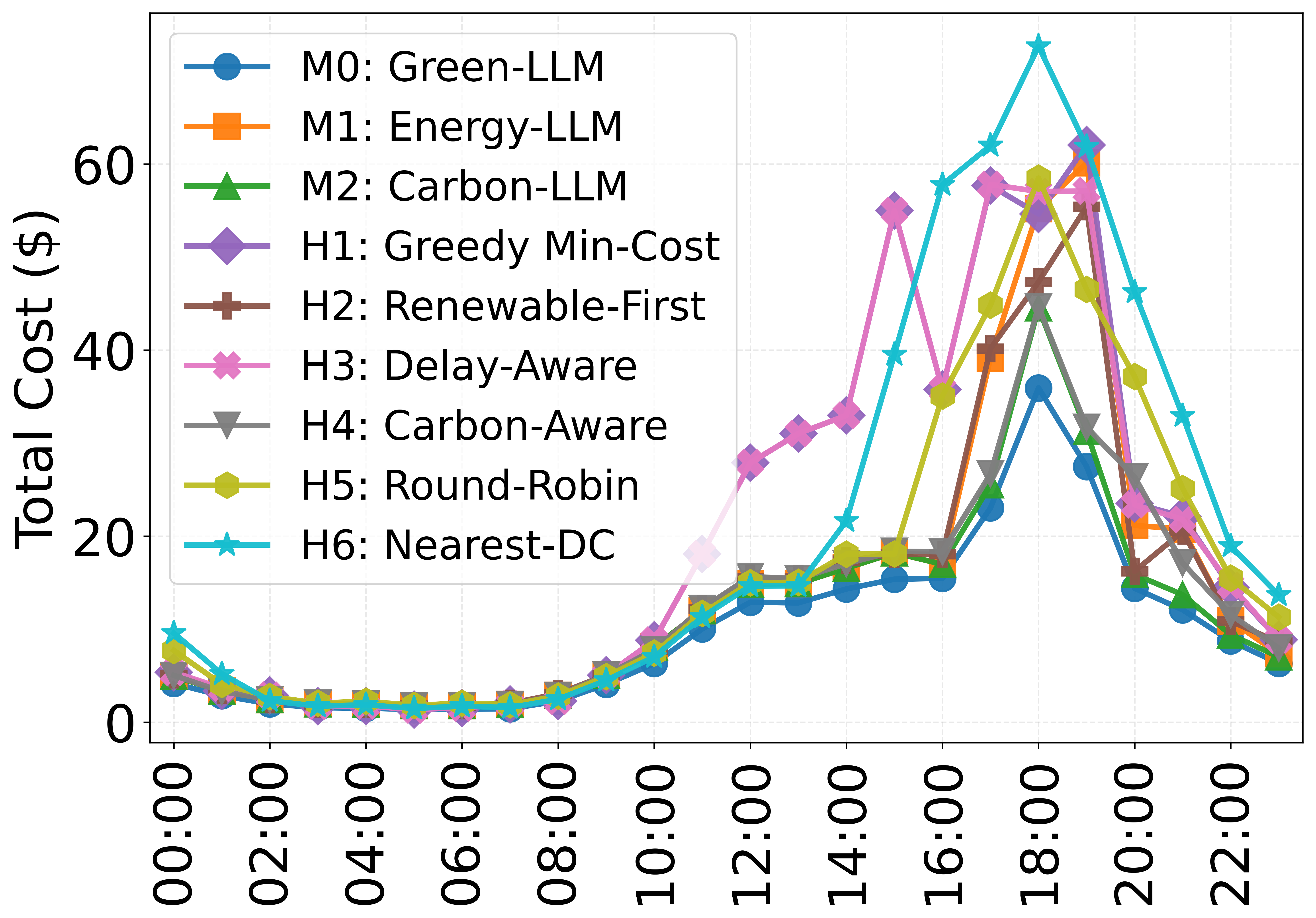}
	     \label{fig:Theta_cost_with_time}
	}  \hspace*{-1em} 
	     \subfigure[Carbon emission (t): $\Psi_\theta = 0.7$]{
	     \includegraphics[width=0.23\textwidth,height=0.125\textheight]{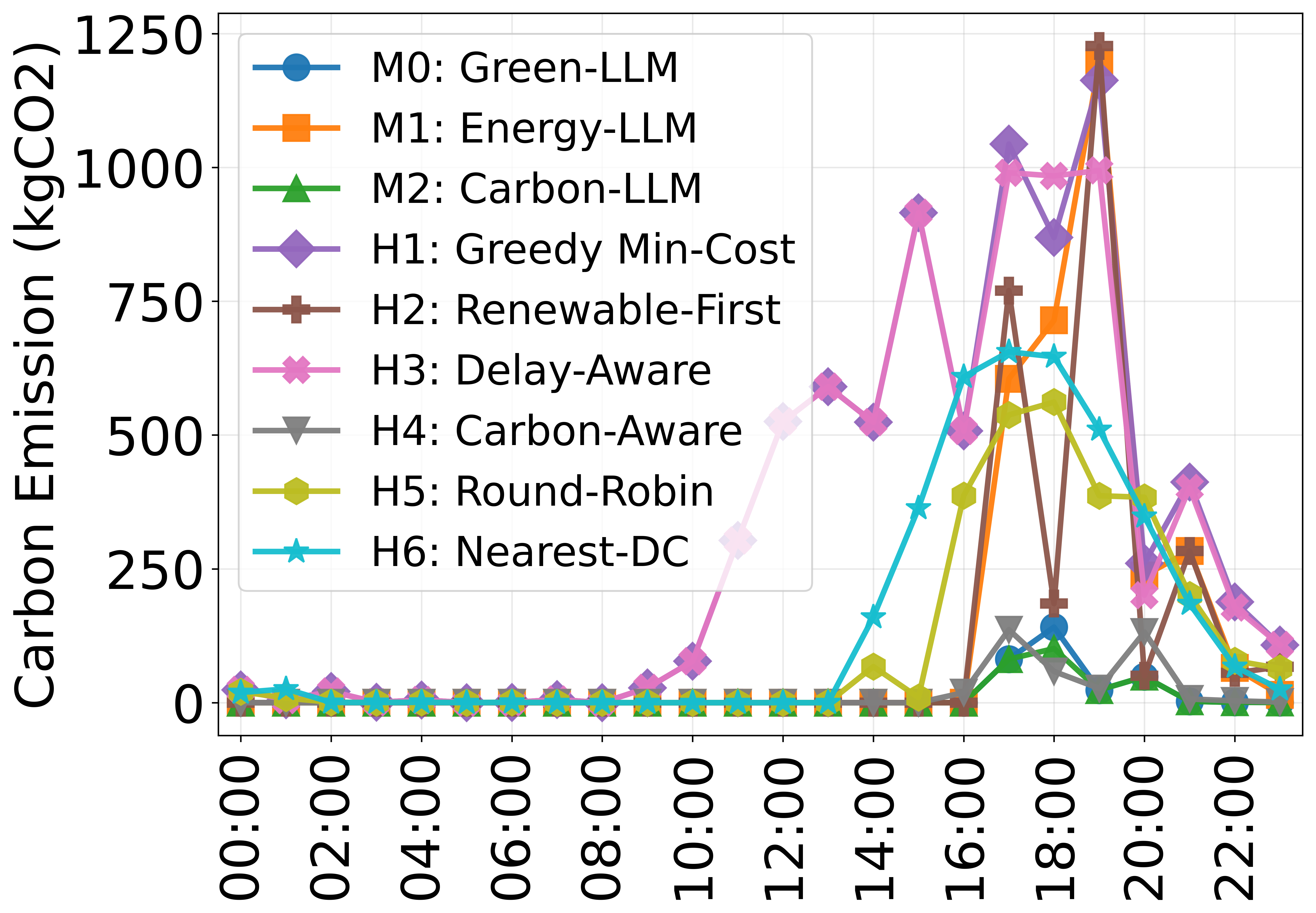}
    	     \label{fig:Theta_carbon_with_time}}\vspace{-0.3cm}
    \subfigure[Carbon emission (t): $\Psi_w = 0.7$]{
	     \includegraphics[width=0.23\textwidth,height=0.125\textheight]{plot_python/Pw_carbon_time_series_final.png}
	     \label{fig:Pw_carbon_with_time}
	}\hspace*{-1em}
	     \subfigure[Delay penalty (t): $\Psi_w = 0.7$]{
	     \includegraphics[width=0.23\textwidth,height=0.125\textheight]{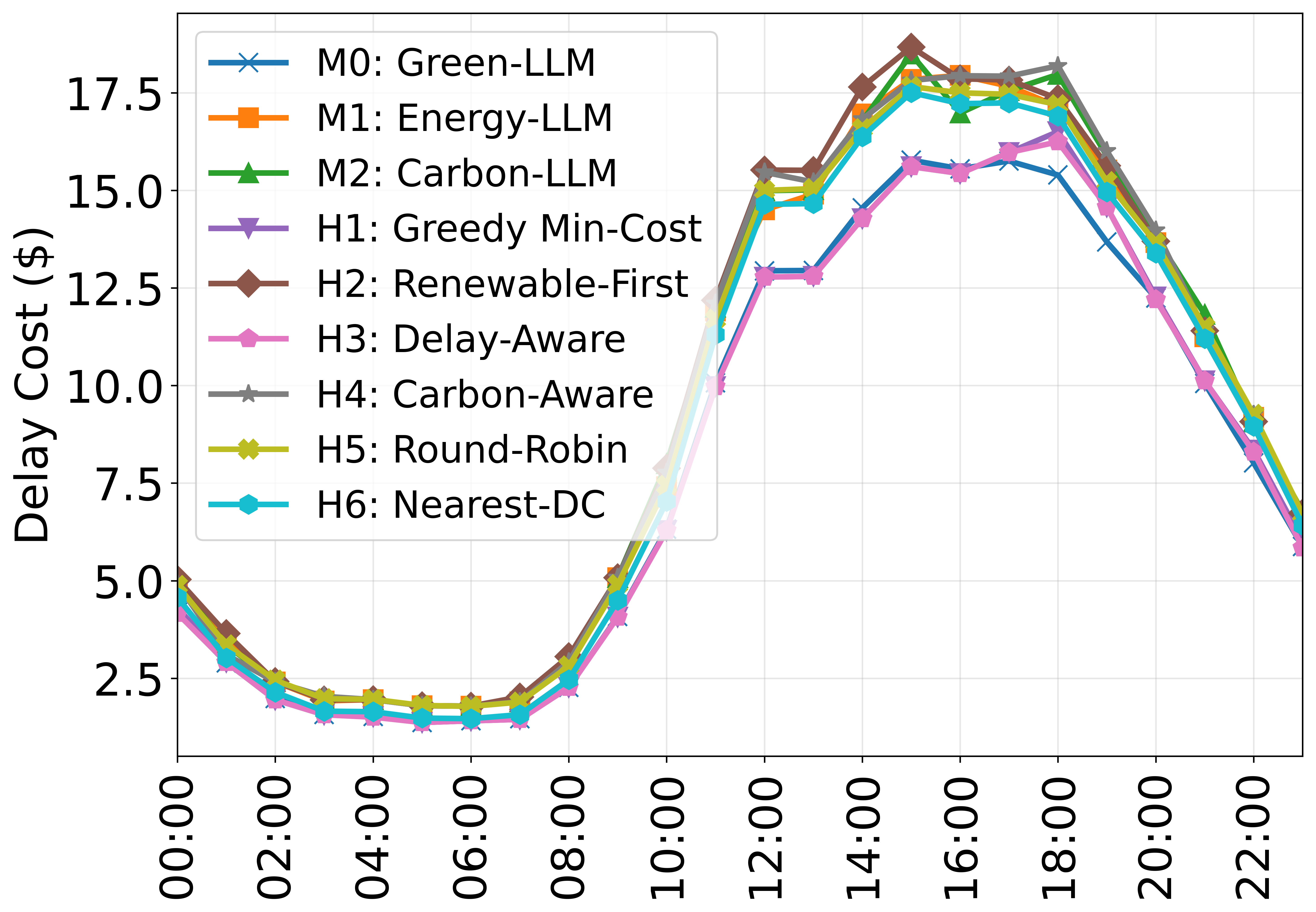}
	     \label{fig:Pw_delay_with_time}}\vspace{-0.2cm}
	    \caption{Varying carbon intensity $\theta$ and renewable energy $P_w$: \textit{effective cost and carbon reduction requires joint, time-aware multi-objective optimization.}}
        \vspace{-0.1cm}
\end{figure}

\noindent \textit{\textbf{1) Varying carbon intensity:}} Fig.~\ref{fig:Theta_cost} shows that \textbf{M0} consistently achieves the lowest cost with minimal sensitivity to $\Psi_\theta$. Carbon-aware approaches (\textbf{M2},\textbf{H4}) show mild cost increases, reflecting their preference for low-carbon DCs. In contrast, energy and delay-driven heuristics (\textbf{H1}--\textbf{H3},\textbf{H5}--\textbf{H6}) experience steep, linear cost growth as $\Psi_\theta$ rises, since they fail to internalize time-varying carbon intensity. Figs.~\ref{fig:Theta_cost_with_time}--\ref{fig:Theta_carbon_with_time} reveal that this divergence is driven by late afternoon and evening emission spikes, when delay-aware heuristics are locked into carbon-intensive proximate DCs during peak hours.

\noindent \textit{\textbf{2) Varying renewable energy:}} Fig.~\ref{fig:Pw_cost} shows total cost decreasing with renewable availability, as on-site generation reduces reliance on the carbon-intensive grid. Figs.~\ref{fig:Pw_carbon_with_time}--\ref{fig:Pw_delay_with_time} show that renewable energy delivers the greatest marginal benefit during peak-demand periods, when both delay and carbon costs are high. \textbf{M0} coordinates workload across time and locations to exploit renewable resources during these windows, suppressing congestion and carbon peaks. \textbf{M1} and \textbf{M2} improve over heuristics but remain suboptimal due to single-objective myopic decisions.

\smallskip\noindent\textit{\textbf{Key Takeaway 1:}} Joint multi-objective optimization ensures robust cost control across carbon intensities and renewable availability; its advantage over heuristics peaks during hours requiring high temporal coordination.

\begin{figure}[h!]
\centering
		\subfigure[Total cost: varying $\tau$]{
	     \includegraphics[width=0.23\textwidth,height=0.125\textheight]{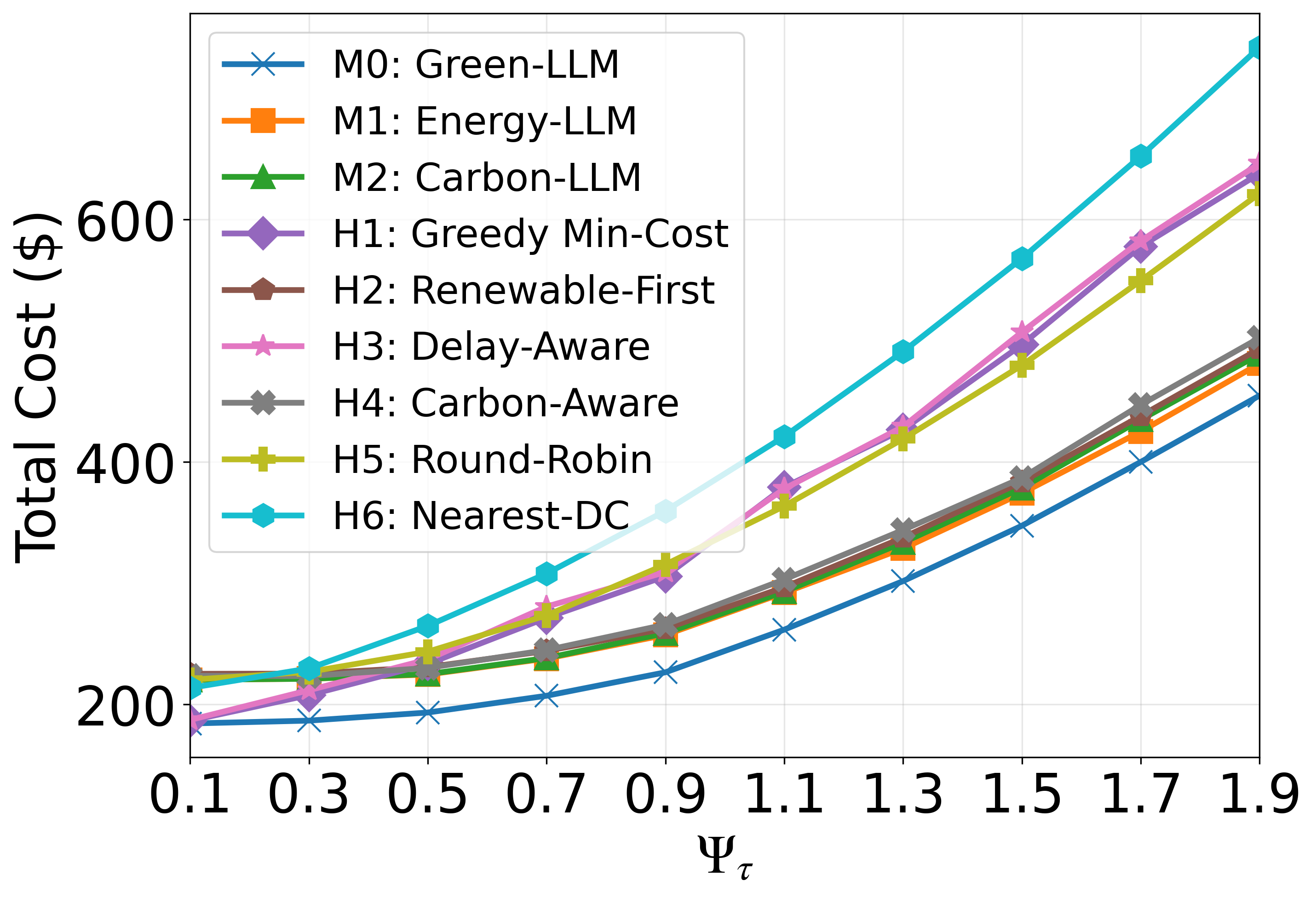}
	     \label{fig:Tau_cost}
	}  \hspace*{-1.5em} 
	     \subfigure[Carbon emission: varying $\tau$]{
	     \includegraphics[width=0.23\textwidth,height=0.125\textheight]{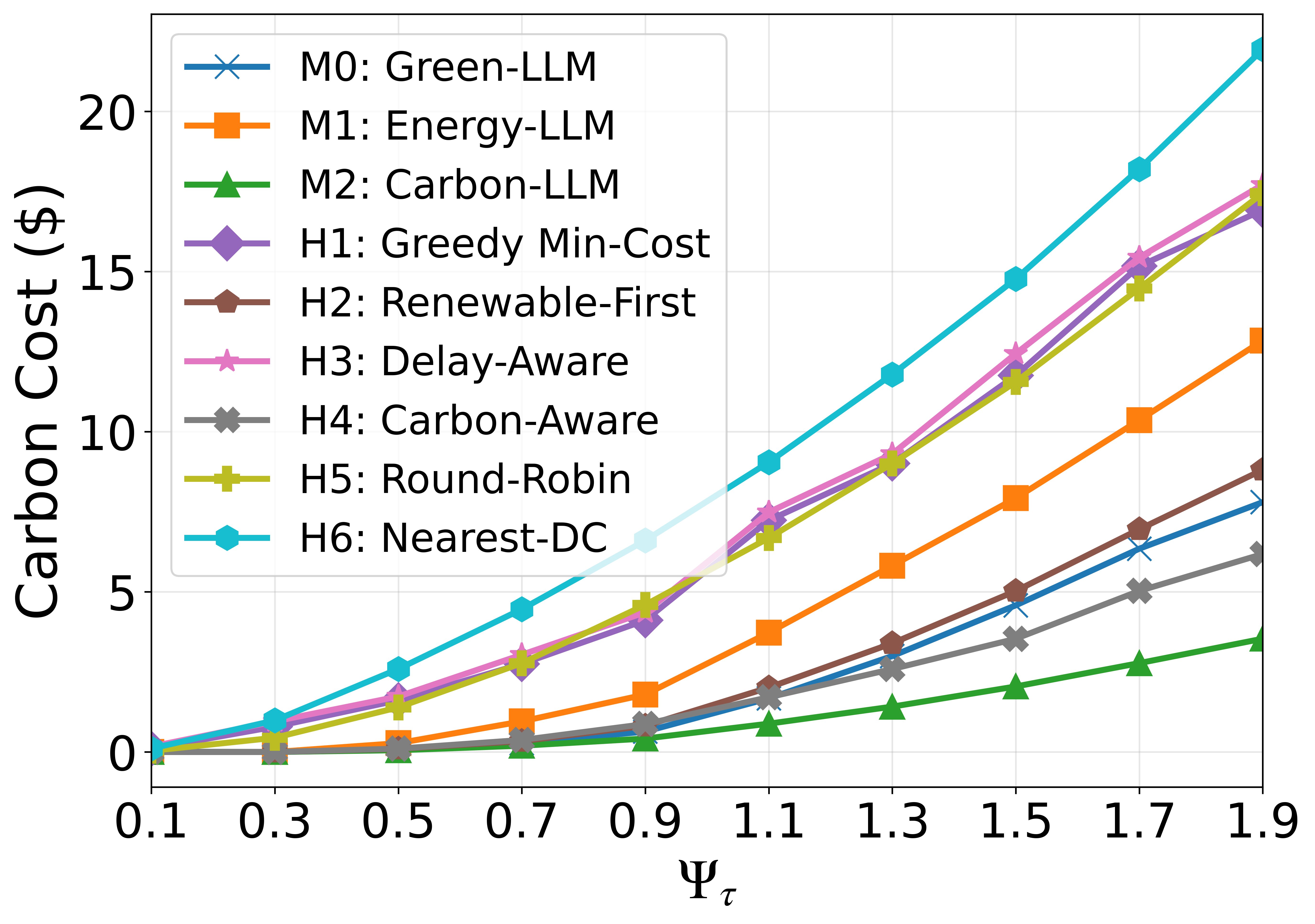}
	     \label{fig:Tau_carbon}
	}\vspace{-0.2cm}
    \subfigure[Delay: varying $\rho$]{
	     \includegraphics[width=0.23\textwidth,height=0.125\textheight]{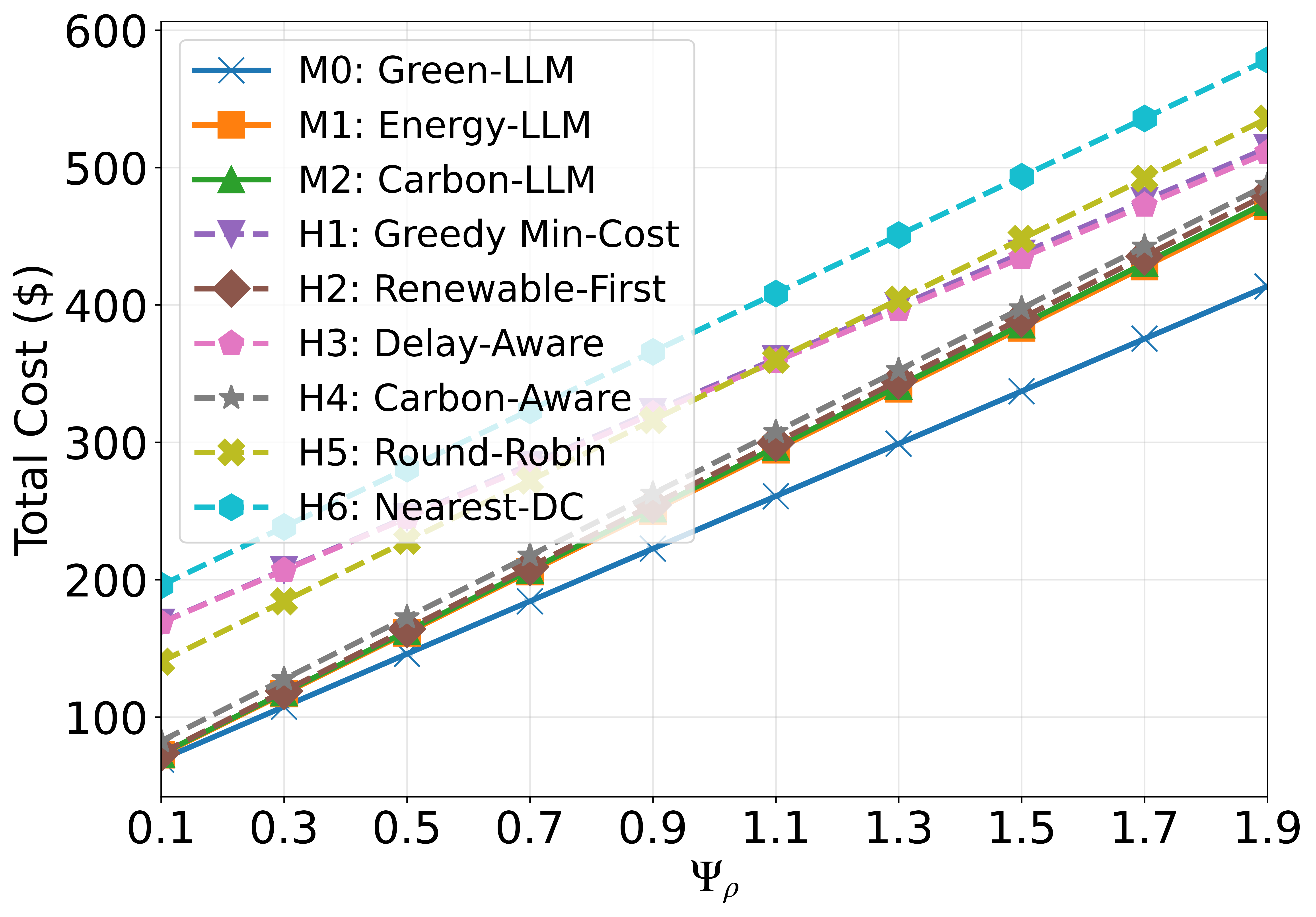}
	     \label{fig:rho_delay}
	} \hspace*{-1.5em} 
		\subfigure[Total cost: varying $\rho$ and $\tau$]{
	     \includegraphics[width=0.23\textwidth,height=0.125\textheight]{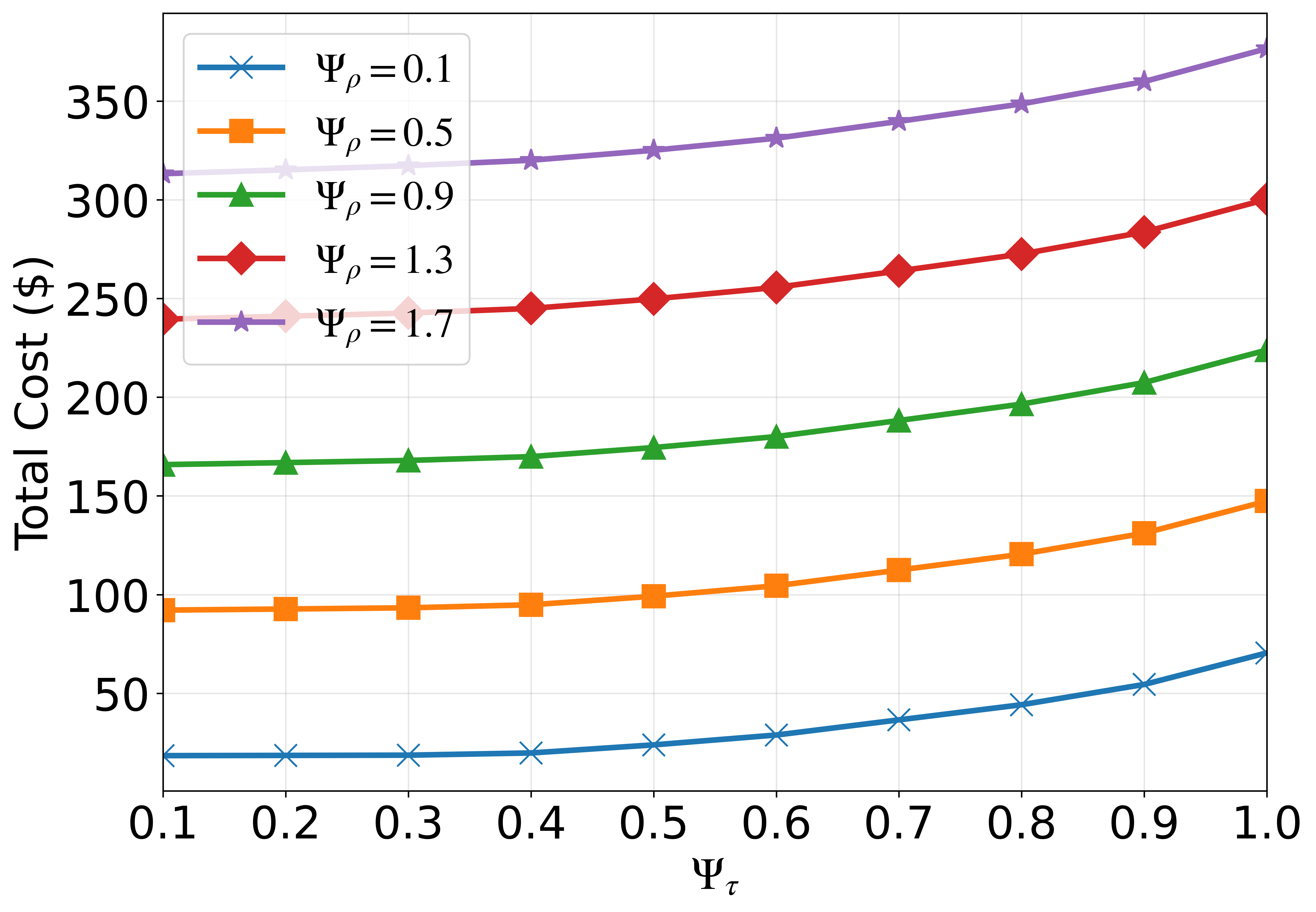}
	     \label{fig:tau_rho_cost}
	} \vspace{-0.2cm}
    \caption{Varying energy coefficient $\tau$ and delay penalty $\rho$: \textit{M0 maintains low cost when high per-token energy and delay penalties coexist.}}
    \vspace{-0.3cm}
\end{figure}

\noindent \textit{\textbf{3) Impacts of other parameters:}} Figs.~\ref{fig:Tau_cost}-\ref{fig:Tau_carbon} show that increasing the energy coefficient scaling $\Psi_\tau$ raises cost and carbon sharply for all models, as larger tokens correspond to longer processing durations and higher computational load. \textbf{M1} is most sensitive to token size, while \textbf{M2} maintains low emissions at a higher cost; \textbf{M0} achieves a favorable balance. Heuristics exhibit the steepest growth, as single-metric method cannot adapt to the compounded effects of computation intensity and congestion. Fig.~\ref{fig:rho_delay} confirms a similar pattern for delay penalties $\Psi_\rho$, with heuristics becoming increasingly unstable when large token sizes amplify delay-induced penalties.

\smallskip\noindent\textit{\textbf{Key Takeaway 2:}} As LLM workloads become more compute-intensive, adaptive multi-objective optimization becomes increasingly critical; static heuristics degrade rapidly under compounding resource pressures.

\noindent \textit{\textbf{4) Method selections:}}
Tables~\ref{tab:lexicographic}--\ref{tab:weighted_sum} compare lexicographic optimizaiton and weighted-sum approach. For example, even a moderate increase in the carbon weight (such as $\sigma_c = 0.6$) can cut carbon emissions by almost half while increasing total cost by less than $1\%$. In contrast, the lexicographic approach strictly follows a preset priority order, which can lead to much larger changes in performance. Based on our results, total cost can more than double when the highest-priority objective changes. This makes lexicographic optimization especially useful in settings with strict policy or regulatory requirements, where one objective must be satisfied first before improving others. For instance, if carbon performance must meet a required limit before cost is considered, the lexicographic approach guarantees that ordering by design, whereas a weighted formulation cannot provide the same guarantee with a single fixed set of weights.

\smallskip\noindent\textit{\textbf{Key Takeaway 3:}} Weighted-sum and lexicographic formulations are complementary: the former enables fine-grained trade-off exploration during planning, while the latter guarantees strict priority compliance required by regulatory or contractual constraints.

\begin{table}[t]
\centering
\caption{Lexicographic optimization results under different priority orders (E: Energy, C: Carbon, D: Delay). }
\label{tab:lexicographic}
\begin{tabular}{|c|c|c|c|c|}
\hline
\textbf{\begin{tabular}[c]{@{}c@{}}Priority \\ Order\end{tabular}} & \textbf{\begin{tabular}[c]{@{}c@{}}Total \\ Cost (\$)\end{tabular}} & \textbf{\begin{tabular}[c]{@{}c@{}}Energy \\ Cost (\$)\end{tabular}} & \textbf{\begin{tabular}[c]{@{}c@{}}Carbon \\ Cost (\$)\end{tabular}} & \textbf{\begin{tabular}[c]{@{}c@{}}Delay \\ Penalty (\$)\end{tabular}} \\ \hline
E $\succ$ D $\succ$ C                                              & 394.46                                                         & 193.94                                                          & 8.14                                                                & 192.19                                                            \\ \hline
E $\succ$ C $\succ$ D                               & 395.66                                                         & 194.13                                                          & 7.84                                                                & 193.68                                                            \\ \hline
D $\succ$ E $\succ$ C                                              & 628.24                                                         & 425.72                                                          & 22.51                                                               & 180.00                                                            \\ \hline
D $\succ$ C $\succ$ E                                              & 642.20                                                         & 443.58                                                          & 18.62                                                               & 180.00                                                            \\ \hline
C $\succ$ E $\succ$ D                                              & 404.72                                                         & 207.13                                                          & 2.09                                                                & 195.49                                                            \\ \hline
C $\succ$ D $\succ$ E                                              & 404.89                                                         & 207.11                                                          & 2.09                                                                & 195.69                                                            \\ \hline
\end{tabular}
\end{table}

\begin{table}[t]
\centering
\caption{Weighted-sum optimization results.}
\label{tab:weighted_sum}
\begin{tabular}{|c|c|c|c|c|}
\hline
\textbf{\begin{tabular}[c]{@{}c@{}}Weight Vector\\ ($\sigma_e,\sigma_c,\sigma_d$)\end{tabular}} & \textbf{\begin{tabular}[c]{@{}c@{}}Total \\ Cost (\$)\end{tabular}} & \textbf{\begin{tabular}[c]{@{}c@{}}Energy\\ Cost (\$)\end{tabular}} & \textbf{\begin{tabular}[c]{@{}c@{}}Carbon\\ Cost (\$)\end{tabular}} & \textbf{\begin{tabular}[c]{@{}c@{}}Delay\\ Penalty (\$)\end{tabular}} \\ \hline
(0.33, 0.33, 0.33)                                                                              & 392.08                                                         & 197.63                                                         & 5.46                                                           & 189.00                                                           \\ \hline
(0.60, 0.20, 0.20)                                                                              & 394.01                                                         & 194.32                                                        & 8.01                                                           & 191.68                                                           \\ \hline
(0.20, 0.60, 0.20)                                                                              & 393.89                                                         & 202.86                                                         & 2.96                                                           & 188.08                                                           \\ \hline
(0.20, 0.20, 0.60)                                                                              & 393.41                                                         & 201.64                                                         & 4.55                                                           & 187.23                                                           \\ \hline
(0.45, 0.45, 0.10)                                                                              & 393.42                                                         & 195.33                                                         & 6.29                                                           & 191.80                                                           \\ \hline
(0.45, 0.10, 0.45)                                                                              & 393.13                                                         & 195.64                                                         & 7.64                                                           & 189.85                                                           \\ \hline
\end{tabular}
\vspace{-0.2cm}
\end{table}

\begin{figure}[h!]
\centering
		\subfigure[Total cost vs $\epsilon$]{
	     \includegraphics[width=0.23\textwidth,height=0.125\textheight]{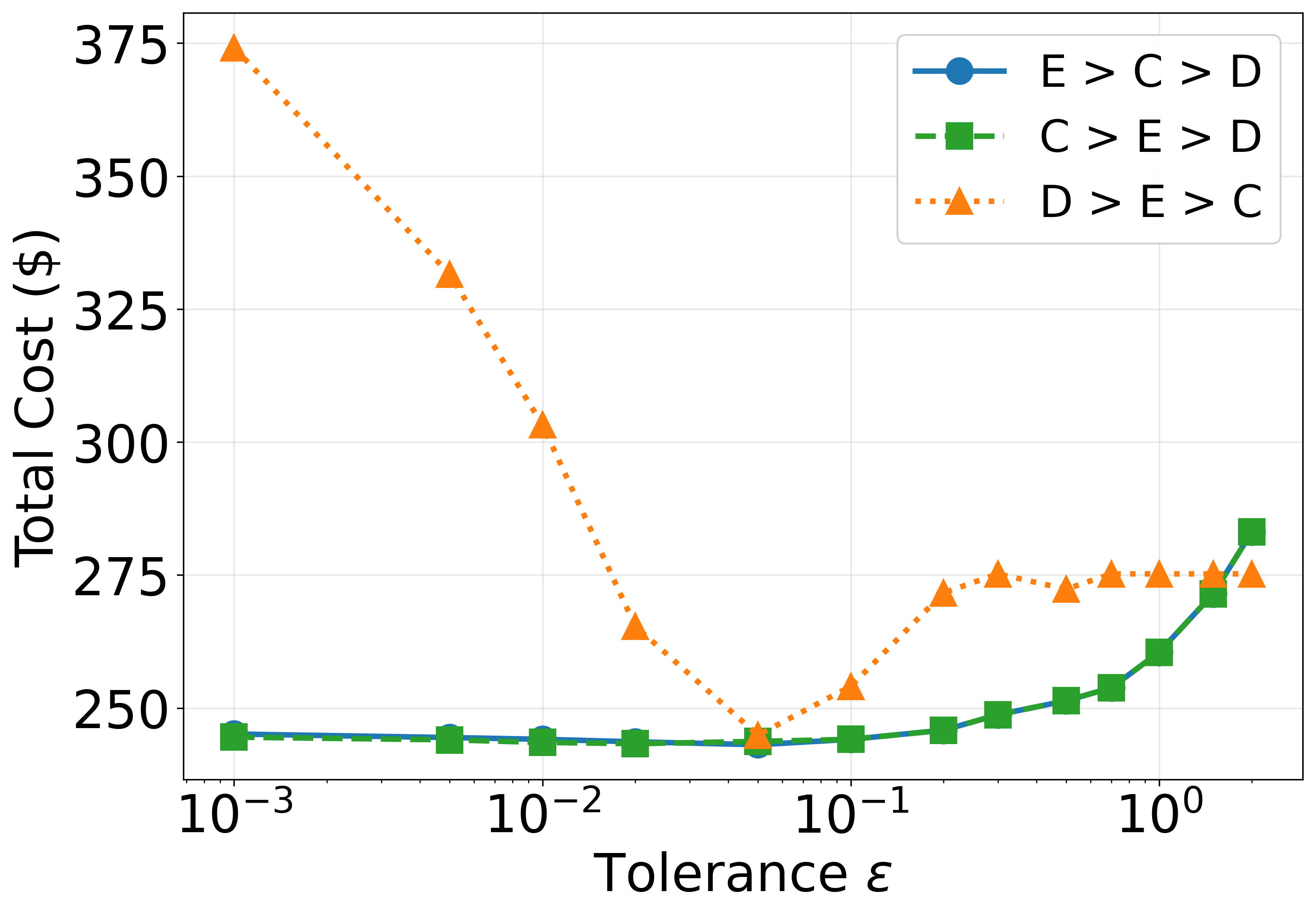}
	     \label{fig:eps_total_cost}
	}  \hspace*{-1em}
	     \subfigure[Cost breakdown: D$\succ$E$\succ$C]{
	     \includegraphics[width=0.23\textwidth,height=0.125\textheight]{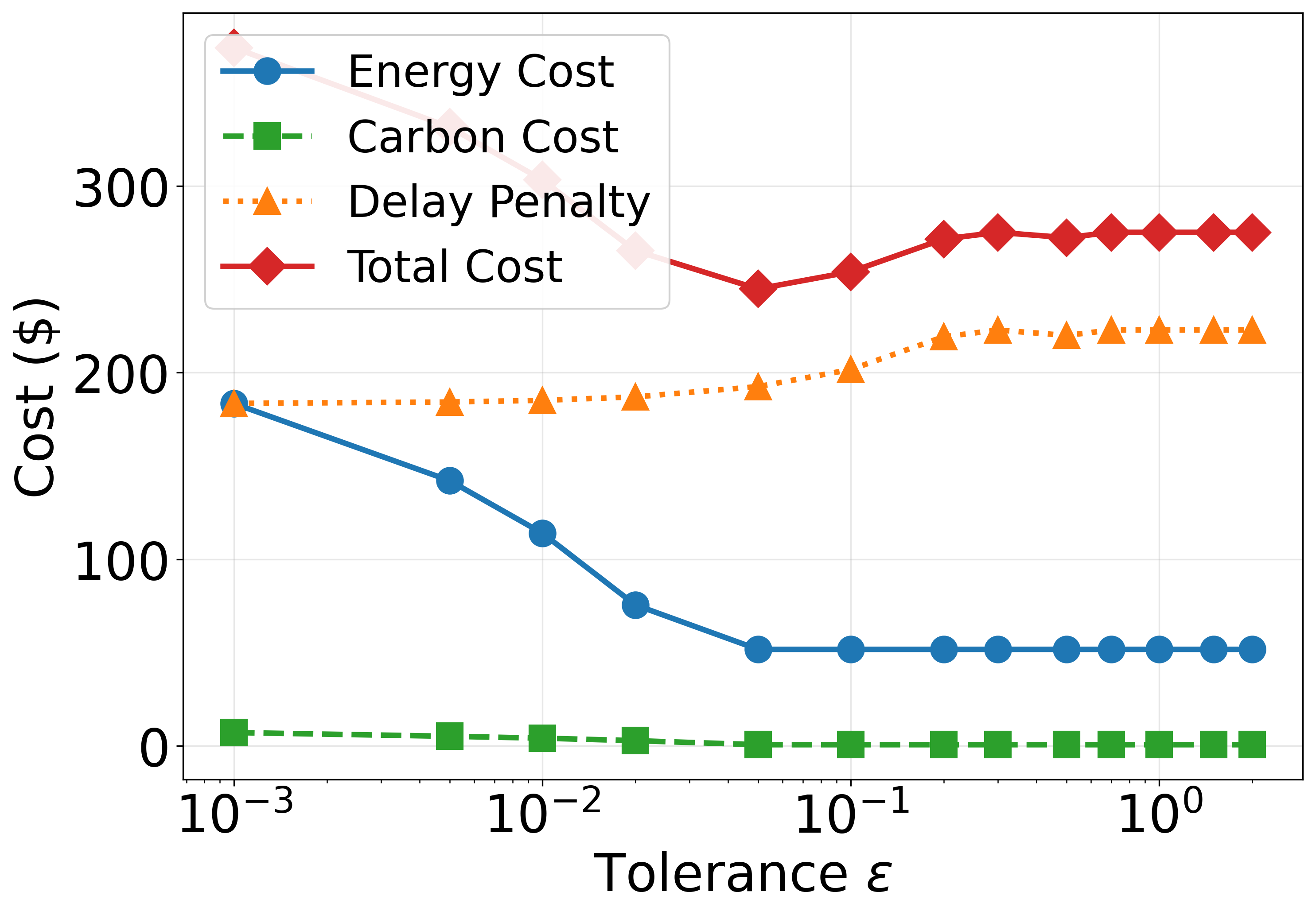}
	     \label{fig:eps_breakdown_dec}
	}\vspace{-0.2cm}
    \caption{Epsilon sensitivity in lexicographic optimization: \textit{delay-first priority is highly sensitive to $\epsilon$ due to geographic lock-in; moderate $\epsilon \approx 0.05$ balances all priorities near the cost-efficient frontier.}}
    \label{fig:epsilon_sensitivity}
    \vspace{-0.3cm}
\end{figure}

\noindent \textit{\textbf{5) Epsilon sensitivity in lexicographic optimization:}}
The tolerance $\epsilon$ in Algorithm~\ref{alg:lexi} controls how much each higher-priority objective may degrade when optimizing lower-priority ones. Fig.~\ref{fig:epsilon_sensitivity} varies $\epsilon \in [0.001, 2.0]$ under three priority orders.

As shown in Fig.~\ref{fig:eps_total_cost}, the sensitivity to $\epsilon$ depends critically on the priority order. Energy- and carbon-first priorities remain stable, because the optimizer retains geographic flexibility to balance grid procurement regardless of the relaxation level. In contrast, the delay-first priority (D$\succ$E$\succ$C) exhibits a U-shaped response: tight $\epsilon$ inflates total cost by over $50\%$, moderate $\epsilon \approx 0.05$ recovers near-optimal cost, and excessive relaxation degrades cost again as workloads spread to distant DCs. This asymmetry stems from a \textit{geographic lock-in} effect: strict delay requirements push workloads into nearby DCs, regardless of electricity price, sacrificing the spatial flexibility that drives cost efficiency.

Fig.~\ref{fig:eps_breakdown_dec} decomposes this behavior for D$\succ$E$\succ$C. Three regimes emerge: (i) at tight $\epsilon < 0.02$, energy cost dominates because the delay constraint confines workloads to expensive proximate DCs; (ii) at moderate $\epsilon \approx 0.05$, a small delay relaxation (${\sim}5\%$) unlocks a disproportionate energy reduction (${\sim}72\%$), yielding the cost minimum; and (iii) beyond $\epsilon > 0.1$, the delay constraint becomes non-binding, energy stabilizes, but delay itself rises as workloads migrate to cheaper, distant locations. The optimal operating point thus lies where the energy--delay trade-off is most asymmetric.

\smallskip\noindent\textit{\textbf{Key Takeaway 4:}} Delay-first lexicographic priority is highly sensitive to $\epsilon$ due to geographic lock-in, while energy- and carbon-first priorities are remain stable. Moderate tolerance ($\epsilon \in [0.01, 0.05]$) resolves this asymmetry, yielding near-optimal total cost regardless of priority ordering.

\noindent \textit{\textbf{6) Water consumption analysis:}}
Although water consumption enters the model as a constraint rather than an objective (i.e., (\ref{constr:water_budget})), workload allocation decisions affect the geographic distribution of water usage across DCs.

\begin{figure}[h!]
\centering
		\subfigure[Cost vs water budget ratio $\Psi_Z$]{
	     \includegraphics[width=0.23\textwidth,height=0.125\textheight]{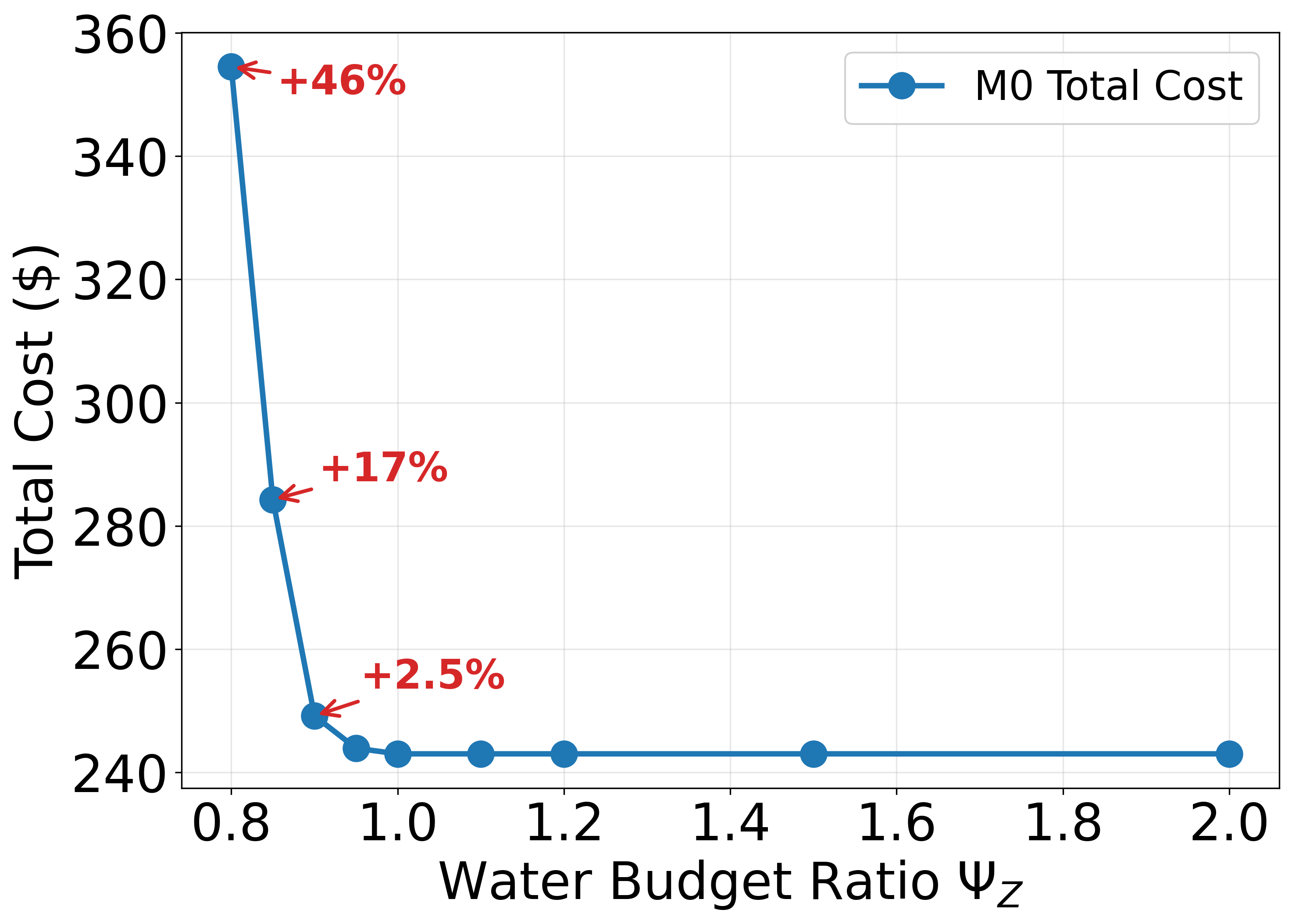}
	     \label{fig:water_stress}
	}  \hspace*{-1em}
	     \subfigure[Per-DC water distribution]{
	     \includegraphics[width=0.23\textwidth,height=0.12\textheight]{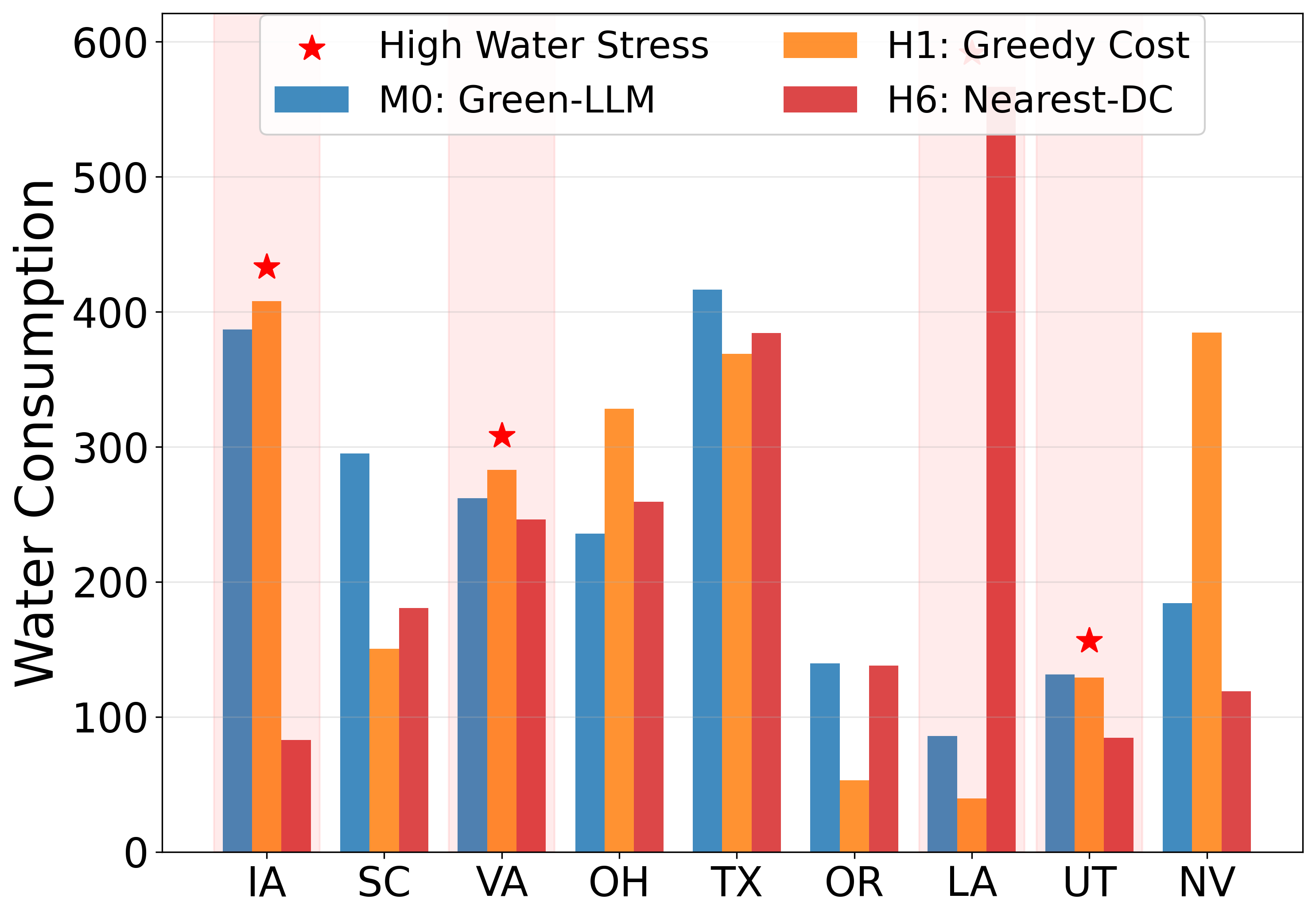}
	     \label{fig:water_per_dc}
	}\vspace{-0.2cm}
    \caption{Water consumption analysis: \textit{tightening the water budget triggers a sharp cost phase transition, and M0 redistributes water away from high-stress regions unlike heuristic methods.}}
    \label{fig:water_analysis}
    \vspace{-0.3cm}
\end{figure}

Fig.~\ref{fig:water_stress} examines the cost of tightening the water budget. We define the budget ratio $\Psi_Z = Z / W_0$, where $W_0$ is the baseline water consumption when the constraint is non-binding. When $\Psi_Z \geq 1.0$, the constraint is slack and cost is unaffected. As $\Psi_Z$ drops below $1.0$, a sharp phase transition emerges: cost rises by only $2.5\%$ at $\Psi_Z = 0.9$ but jumps by $46\%$ at $\Psi_Z = 0.8$, as the optimizer is forced away from cost-efficient DCs to meet the tighter water budget. This nonlinear sensitivity underscores the importance of accurate water budgeting---a modest over-tightening can trigger disproportionate cost penalties.

Fig.~\ref{fig:water_per_dc} compares the per-DC water footprint of \textbf{M0} against two best heuristics with distinct allocation biases: \textbf{H1} (greedy cost) routes to cheapest-electricity DCs, while \textbf{H6} (nearest-DC) routes to geographically closest DCs. Although total water consumption is nearly identical across methods, the geographic distribution differs substantially. \textbf{H6} concentrates $6.6\times$ more water at the LA DC (a high water-stress region) than \textbf{M0}, because proximity-based allocation concentrates Southern California demand in the nearest facility, \textbf{M0} instead shifts part of this load to lower-stress regions (e.g., TX, OR). This leads to a more balanced distribution without increasing total cost. The effect follows naturally from the constrained optimization: the global water budget discourages allocations in stressed regions when comparable alternatives are available.

\smallskip\noindent\textit{\textbf{Key Takeaway 5:}} Tightening the water budget exhibits a sharp cost phase transition below $\Psi_Z = 0.9$; constrained optimization implicitly promotes regional water equity by redistributing workloads away from stressed DCs, a benefit absent in heuristic approaches.

\begin{figure}[h!]
\centering
		\subfigure[Varying ($I,J$)]{
	     \includegraphics[width=0.23\textwidth,height=0.125\textheight]{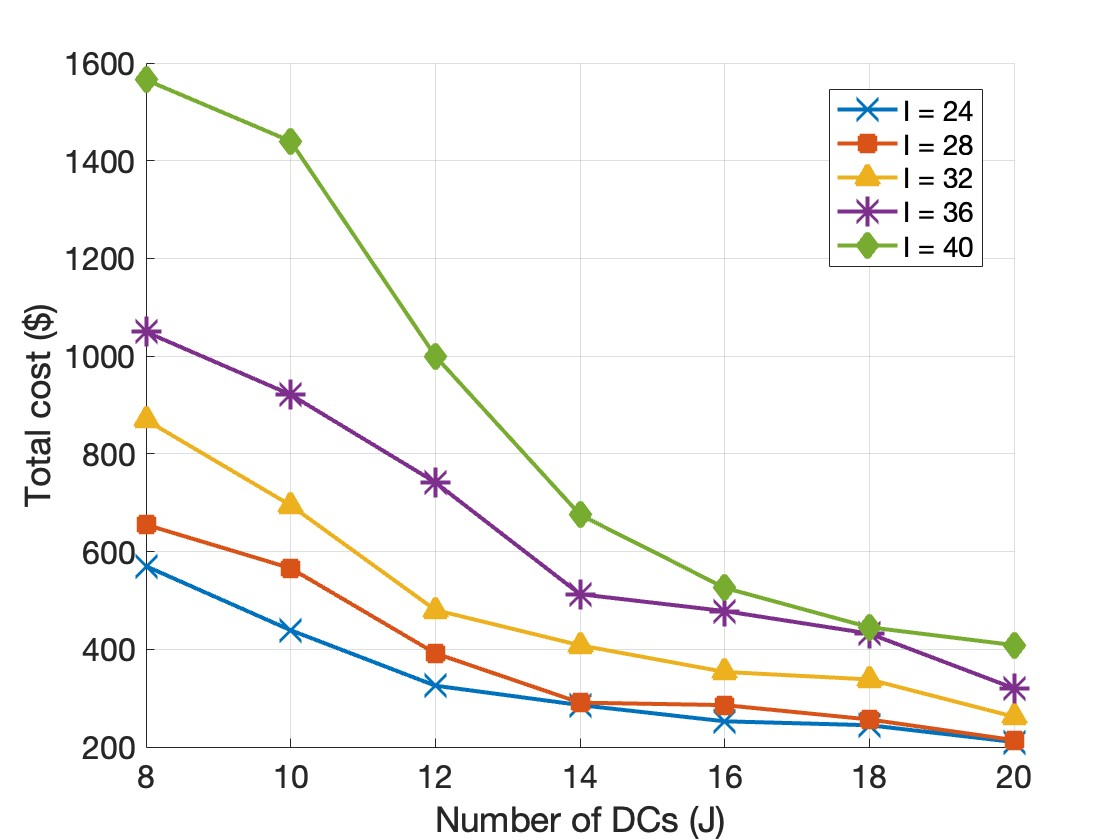}
	     \label{fig:network_size}
	}  \hspace*{-1.5em}
	     \subfigure[Carbon intensity scenarios]{
	     \includegraphics[width=0.23\textwidth,height=0.12\textheight]{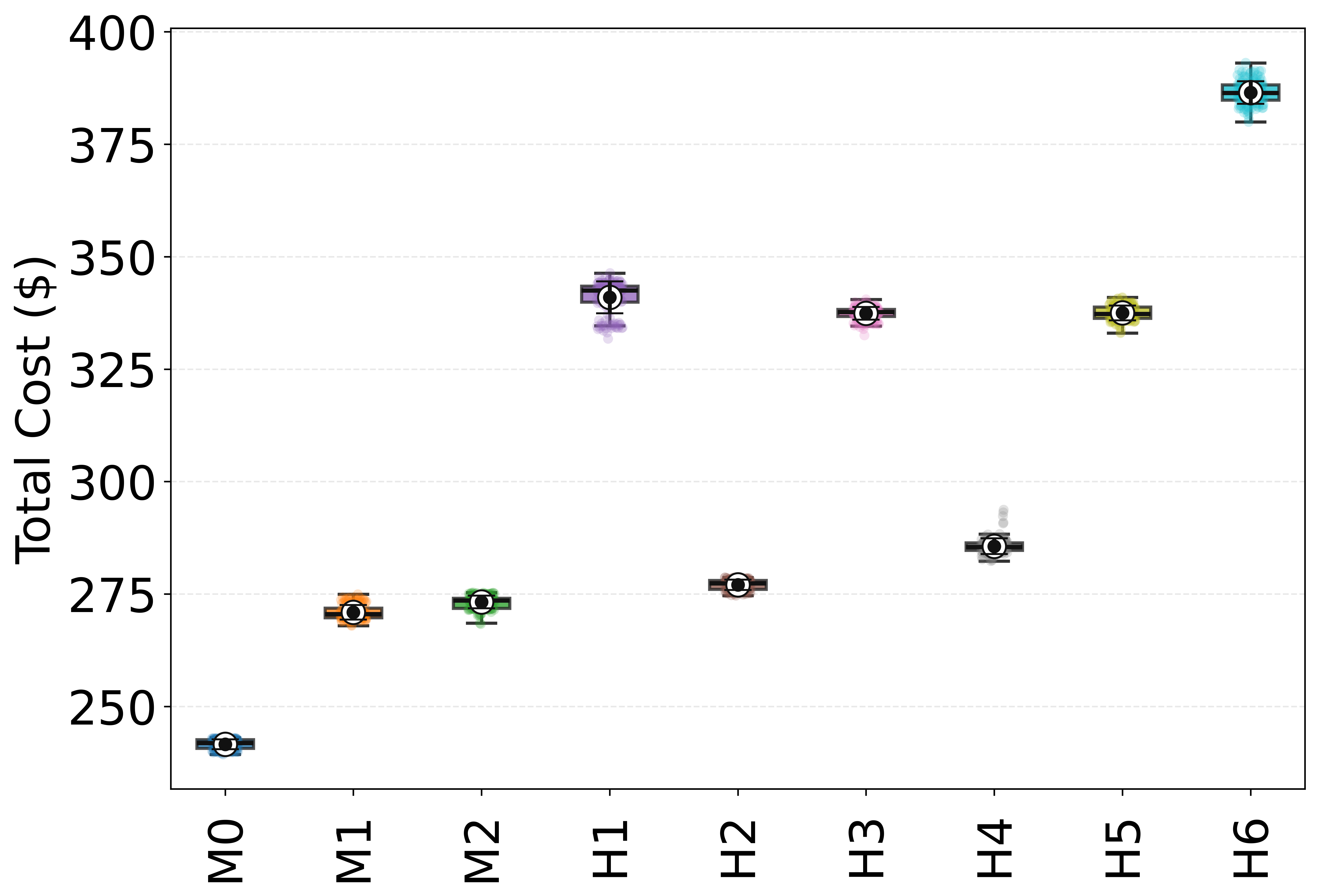}
	     \label{fig:carbon_scenario}
	}
    \caption{Varying network size and robustness evaluation}
    \vspace{-0.2cm}
\end{figure}

\noindent \textit{\textbf{7) Robustness and scalability:}}
We evaluate robustness using $50$ carbon-intensity scenarios derived from real-world data (varying weather, renewable availability, and seasonal patterns), each tested with five random seeds. As shown in Fig.~\ref{fig:carbon_scenario}, \textbf{M0} exhibits minimal cost variation across scenarios, while heuristic methods show noticeable variability due to their myopic, instantaneous decisions. This robustness stems from the system-level optimization and temporal coordination inherent in the multi-objective framework. Table~\ref{tab:running_time} reports computational efficiency across varying network sizes: for a medium-scale instance with $(20,40)$ regions and DCs, the framework solves within $2$ seconds per time period---well below typical decision intervals (e.g., hourly)---demonstrating computational tractability for real-time deployment. Fig.~\ref{fig:network_size} shows that total cost decreases as the number of DCs increases due to improved resource availability and geographic flexibility, with diminishing marginal benefits at larger scales. Conversely, increasing the number of serving areas raises costs due to higher aggregate resource demands.

\vspace{-0.2cm}

\smallskip\noindent\textit{\textbf{Key Takeaway 6:}} The optimization framework delivers stable performance across diverse environmental scenarios and remains computationally tractable for real-time deployment.

\begin{table}[t]
\centering
\begin{tabular}{|ccc|c||ccc|c|}
\hline
$I$ & $J$ & Status & Time (s) & $I$ & $J$ & Status & Time (s) \\ \hline
8  & 32 & OPT & 0.62 & 16 & 32 & OPT & 1.03 \\
8  & 36 & OPT & 0.72 & 16 & 36 & OPT & 1.17 \\
8  & 40 & OPT & 0.92 & 16 & 40 & OPT & 1.57 \\ \cline{1-8}
12 & 32 & OPT & 0.71 & 20 & 32 & OPT & 1.25 \\
12 & 36 & OPT & 1.01 & 20 & 36 & OPT & 1.38 \\
12 & 40 & OPT & 1.15 & 20 & 40 & OPT & 1.56 \\ \hline
\end{tabular}
\caption{Run time analysis vs. network sizes}
\label{tab:running_time}
\vspace{-0.5cm}
\end{table}

\section{Conclusion}
\label{sec:conclusion}
This paper introduces Green-LLM, a multi-objective framework for sustainable LLM inference across heterogeneous edge data centers. By jointly minimizing energy costs, carbon emissions, and delay penalties under strict water constraints, the framework utilizes a lexicographic optimization approach ensuring polynomial-time complexity for real-time deployment without manual weight tuning. Sensitivity analysis of the tolerance parameter $\epsilon$ indicates that moderate values provide an optimal balance between strict priority compliance and total cost efficiency. Furthermore, our analysis reveals that Green-LLM promotes regional water equity through geographic load balancing, avoiding the resource strain typical of proximity-driven heuristics. Numerical results demonstrate significant environmental savings while maintaining operational costs within 3\% of the minimum and ensuring sub-2-second end-to-end latency. Future work will explore uncertainty-aware extensions and joint optimization of model placement with workload allocation.


\end{document}